\documentclass[sigconf, 9pt]{acmart}

\usepackage{url}
\usepackage{graphicx}
\usepackage{algorithmic}
\usepackage{subfigure}
\usepackage{amsmath}
\usepackage{xcolor}
\usepackage{makecell}
\usepackage{booktabs}
\usepackage{multirow}
\usepackage{balance}
\usepackage{listings}
\usepackage{booktabs}
\usepackage{pifont}
\usepackage{enumitem}
\usepackage[ruled,vlined,linesnumbered]{algorithm2e}
\usepackage[inkscapelatex=false]{svg}
\usepackage{bm}
\usepackage{array}
\usepackage{ragged2e}
\usepackage{tabularray}
\usepackage[normalem]{ulem}
\usepackage{enumitem}
\newcommand{\project}{\textit{GPIoT} }
\newcommand{\ie}{\textit{i}.\textit{e}.}
\newcommand{\eg}{\textit{e}.\textit{g}.}
\newcommand{\etc}{\textit{etc}}

\begin{document}
\begin{sloppypar}

\title{GPIoT: Tailoring Small Language Models for IoT Program Synthesis and Development}

\author{Leming Shen\textsuperscript{1},
    Qiang Yang\textsuperscript{2},
    Xinyu Huang\textsuperscript{1},
    Zijing Ma\textsuperscript{1},
    Yuanqing Zheng\textsuperscript{1}
}

\affiliation{
    \textsuperscript{1}The Hong Kong Polytechnic University,
    \textsuperscript{2}University of Cambridge
    \country{}
}

\email{
    {leming.shen, unixy-xinyu.huang, zijing.ma}@connect.polyu.hk, qy258@cam.ac.uk, csyqzheng@comp.polyu.edu.hk,
}

\renewcommand{\shortauthors}{Leming Shen, Qiang Yang, Xinyu Huang, Zijing Ma, Yuanqing Zheng}

\begin{abstract}
Code Large Language Models (LLMs) enhance software development efficiency by automatically generating code and documentation based on user requirements. However, code LLMs cannot synthesize specialized programs when tasked with IoT applications that require domain knowledge. While Retrieval-Augmented Generation (RAG) offers a promising solution by fetching relevant domain knowledge, it necessitates powerful cloud LLMs (\eg, GPT-4) to process user requirements and retrieved contents, which raises significant privacy concerns. This approach also suffers from unstable networks and prohibitive LLM query costs. Moreover, it is challenging to ensure the correctness and relevance of the fetched contents. To address these issues, we propose \textit{GPIoT}, a code generation system for IoT applications by fine-tuning locally deployable Small Language Models (SLMs) on IoT-specialized datasets. SLMs have smaller model sizes, allowing efficient local deployment and execution to mitigate privacy concerns and network uncertainty. Furthermore, by fine-tuning SLMs with our IoT-specialized datasets, the SLMs' ability to synthesize IoT-related programs can be substantially improved. 
To evaluate \textit{GPIoT}'s capability in synthesizing programs for IoT applications, we develop a benchmark, \textit{IoTBench}.
Extensive experiments and user trials demonstrate the effectiveness of \textit{GPIoT} in generating IoT-specialized code, outperforming state-of-the-art code LLMs with an average task accuracy increment of 64.7\% and significant improvements in user satisfaction.

\end{abstract}

\ccsdesc[500]{Computing methodologies~Artificial intelligence}
\ccsdesc[500]{Computer systems organization~Embedded and cyber-physical systems}

\keywords{Small Language Model, IoT Program Synthesis, Fine-tuning}

\acmYear{2025}\copyrightyear{2025}
\acmConference[SenSys '25]{The 23rd ACM Conference on
Embedded Networked Sensor Systems}{May 6--9, 2025}{Irvine, USA}
\acmBooktitle{The 23rd ACM Conference on
Embedded Networked Sensor Systems (SenSys '25), May 6--9, 2025, Irvine, USA}
\acmDOI{XXXXXXX.XXXXXXX}
\acmISBN{978-1-4503-XXXX-X/2018/06}

\maketitle
\vspace{-5pt}
\section{Introduction}
\begin{figure}[t]
\setlength{\abovecaptionskip}{2pt}
    \centering
    \includegraphics[width=0.45\textwidth]{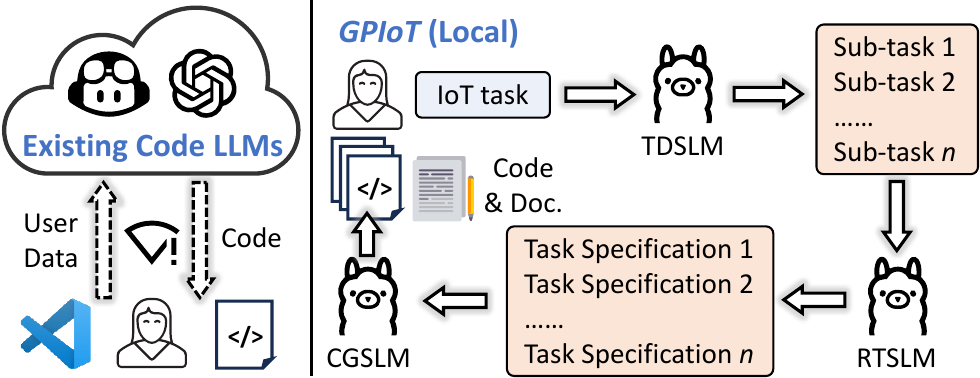}
    \caption{Existing code LLMs need to transmit sensitive data to remote servers. In contrast, \project features three local SLMs to protect user privacy and reduce query costs.}
    \label{fig:copilot}
\vspace{-15pt}
\end{figure}

\vspace{-3pt}
Large language models (LLMs) \cite{wan2023efficient, wang2024svd} are revolutionizing various aspects of embedded system development and mobile computing, \eg, smartphone task automation \cite{wen2024autodroid}, advanced virtual assistants \cite{nie2024llm}, and even IoT data comprehension \cite{xu2024penetrative, ouyang2024llmsense, liu2024tasking, an2024iot}. Code LLMs (\eg, WizardCoder \cite{luo2023wizardcoder} and CodeLlama \cite{roziere2023code}) stand out as promising tools designed to synthesize programs based on user requirements described in natural language. As illustrated in Fig.~\ref{fig:copilot}, the integration of programming tools with code LLMs significantly enhances software development by automating code completion, code generation, bug detection, documentation writing, \etc.

While powerful and promising, when confronted with IoT applications \cite{cao2024practical, yu2024revolutionizing, yu2024fdlora, yang2023aquahelper, yang2024neural, yang2021model, hou2024one, hou2023don} that require special domain knowledge, existing code LLMs tend to simply provide general solutions with sub-optimal performance  (\S~\ref{s:preliminary}).
This is because they focus on general-purpose programming tasks \cite{jiang2024survey} rather than being tailored to any particular domain. Moreover, IoT-related knowledge and programs only occupy a small proportion of the datasets which code LLMs were trained on \cite{zhao2023survey}. Consequently, IoT terminologies will be assigned a lower priority during inference with the generated code less dedicated to the IoT domain (\S~\ref{s:preliminary}). This motivates the following research question: \textit{Can we build a code LLM specially tailored for IoT application code generation tasks?} If yes, we can synthesize IoT-related programs with higher task accuracy and efficiency, offering significant convenience for IoT developers.



A potential approach can be Retrieval-Augmented Generation (RAG) \cite{lewis2020retrieval}, which provides LLMs with retrieved domain knowledge to enhance their abilities in generating accurate and contextually relevant solutions. Existing works \cite{shen2025autoiot, huang2023agentcoder, ding2024reasoning} construct a sophisticated LLM+RAG agent to gradually generate code through multiple intermediate steps via prompts. Nonetheless, they suffer from three main problems.
1) A powerful LLM with strong language comprehension capability is needed to learn from the retrieved knowledge. However, cloud LLMs (\eg, GPT-4 \cite{achiam2023gpt}) may suffer from bad network conditions, high costs, and privacy concerns, while local LLMs (\eg, Llama2-70b \cite{touvron2023llama}) have harsh requirements in system resources (\eg, memory and network). 2) Complicated RAG designs (\eg, iterative retrieval \cite{xiong2024improving}) are mandatory to ensure the correctness and high relevance of the retrieved knowledge, with extended processing time. Otherwise, LLMs may fail to focus on the IoT context and still provide general solutions \cite{wu2024faithful}. 3) Meticulously designed prompts are required to ensure that outputs must strictly follow pre-defined formats \cite{lin2024parrot}, which is extremely challenging due to the hallucinations and unreliability of LLMs \cite{ugare2024improving}. 



To tackle the above problems, we propose \textit{GPIoT}, a code generation system tailored for IoT application development by fine-tuning local small language models\footnote{We consider SLMs as open-source language models that can be locally deployed and operated efficiently on commodity GPUs \cite{ibm_small} (\eg, Llama2-13b with INT8 quantization requires around 16 GB of GPU memory).} (SLMs) on IoT-specialized text-generation datasets.
This approach has the following benefits: 1) The system overhead, privacy leakage, and network instability can be mitigated, as SLMs have smaller sizes and can be locally deployed without incurring heavy resource burdens. 2) SLMs tuned on IoT-specialized datasets can generate responses with significantly enhanced quality and higher relevance to the IoT domain \cite{shen2024iotcoder}. 
3) As our tuning datasets are well-structured text data, the tuned SLMs can produce intermediate outputs following the expected format with enhanced stability and avoid hallucinations.

We implement \textit{GPIoT}\footnote{The models and datasets are available at \url{https://github.com/lemingshen/GPIoT}} with three tailored SLMs to handle different stages of IoT application development: TDSLM for Task Decomposition, RTSLM for Requirement Transformation, and CGSLM for Code Generation. 
Though tuning one SLM to handle all the tasks is possible, it is extremely challenging due to SLMs' limited language understanding and processing capabilities \cite{kaplan2020scaling, lu2024small}. 
As shown in Fig.~\ref{fig:copilot}, TDSLM first decomposes an IoT application into multiple sub-tasks with detailed descriptions. Next, the descriptions are converted into well-structured specifications following pre-defined formats by RTSLM. 
Accordingly, CGSLM further generates a list of code snippets with detailed documentation. By sequentially executing the code based on the documentation, the IoT task can be solved. Note that we only fine-tune TDSLM and CGSLM as these two stages require IoT domain knowledge during inference while RTSLM only needs basic language processing.


In practice, we face three significant technical challenges. 
1) \textbf{Lack of high-quality data.} 
To the best of our knowledge, there are no IoT-oriented user-requirement-to-sub-task and sub-task-to-program text-generation datasets. Thus, we first construct two datasets containing IoT knowledge retrieved from various public sources, aiming to enhance the task decomposition and code generation abilities of TDSLM and CGSLM, respectively. 
Moreover, we design an IoT-oriented text data augmentation method to enhance the datasets' quality and diversity, considering the unique characteristics (\eg, sensor modality and resource heterogeneity) of IoT applications, thereby enhancing SLMs' knowledge comprehension and code generation capabilities for IoT tasks.
2) \textbf{Domain misalignment between SLMs.} 
The decomposed tasks generated by TDSLM may fall beyond the scope that CGSLM can handle due to domain misalignment. This is because the two SLMs focus on different stages of IoT application development during tuning, which could lead to thematic inconsistencies in task interpretation and execution (\S~\ref{s:preliminary}). To tackle this, we propose a parameter-efficient co-tuning (PECT) paradigm featuring a multi-path Low-Rank Adaptation (LoRA) pipeline. 
Unlike conventional LoRA tuning that tunes adapters separately, our designed PECT paradigm enables collaborative fine-tuning of multiple SLMs with a shared base model but with different adapters, thereby mitigating the inconsistency issues and facilitating knowledge sharing between SLMs. 
3) \textbf{Format incompatibility.} Decomposed tasks are typically described in natural language, while the expected inputs of CGSLM should be well-structured. If we directly use the decomposed task descriptions as prompts to generate code, CGSLM can not provide programs strictly following user requirements (\S~\ref{s:preliminary}). To address this, we leverage Chain-of-Thought (CoT) prompting \cite{wei2022chain} that instructs RTSLM to transform the descriptions into well-structured specifications step by step. As such, CGSLM can better handle the specifications to provide IoT-specialized solutions.

To evaluate \textit{GPIoT}, we also propose \textit{IoTBench}, a benchmark to quantify LLMs' capabilities in synthesizing IoT-related programs.
Extensive experiments and a user study demonstrate that \textit{GPIoT} can generate code adopting more IoT-specialized algorithms and outperform SOTA code LLMs in terms of task accuracy (more than 64.7\% on average), memory usage (less than 310 MB on average), and user satisfaction. In summary, we make the following contributions:
\begin{itemize}[leftmargin=9pt]
    \item \textit{GPIoT} presents the first code generation system tailored for IoT application development featuring privacy-preserving local SLMs tuned on IoT-specialized datasets. 
    \item We create IoT domain text-generation datasets with a novel augmentation method tailored for the unique characteristics of IoT tasks, significantly enhancing the IoT knowledge comprehension ability of our tuned SLMs. We also construct \textit{IoTBench} to evaluate the capability of LLMs in synthesizing IoT-specialized programs.
    \item 
    We propose the PECT paradigm, a new LLM tuning method that can collaboratively fine-tune multiple SLMs to mitigate their domain misalignment with facilitated knowledge exchange.
\end{itemize}

\vspace{-5pt}
\section{Background \& Motivation}

\vspace{-2pt}
We first revisit existing code LLMs to underscore the importance of constructing tailored IoT-related LLMs. Then, we conduct some preliminary experiments on existing LLM+RAG methods to further motivate our work with several challenges we need to address.

\vspace{-5pt}
\subsection{Code LLM and LLM+RAG}
\label{s:background}
\begin{figure}[t]
\vspace{-8pt}
\setlength{\abovecaptionskip}{-3pt}
\subfigtopskip=-2pt
\subfigcapskip=-2pt
    \centering
    \subfigure[Terminologies in vector space]{
        \label{fig:word_vector_space}
        \includegraphics[width=0.218\textwidth]{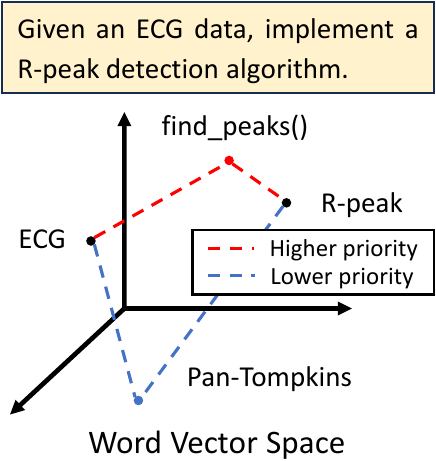}
    }
    \centering
    \subfigure[General workflow of LLM+RAG]{
        \label{fig:autoiot_limitation}
        \includegraphics[width=0.232\textwidth]{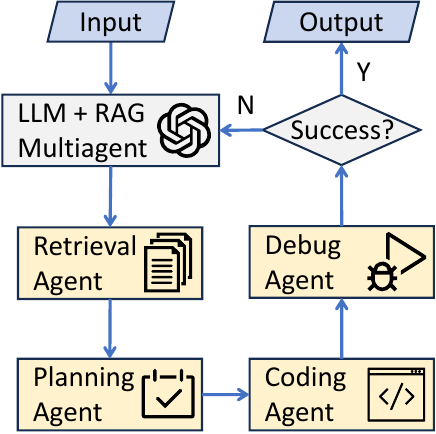}
    }
    \caption{(a) Existing LLMs tend to prioritize general terms; (b) LLM+RAG systems require multiple cascaded agents.}
    \vspace{-15pt}
\end{figure}

\vspace{-2pt}
Existing code LLMs aim to synthesize programs and enhance software development efficiency and accuracy. While they perform well on general and simple programming tasks (\eg, sorting algorithms), they often struggle with complex problems in the IoT domain. For example, when prompted to design an R-peak detection method for electrocardiogram (ECG) data, existing code LLMs can only use the \verb|find_peaks()| function, which adopts a general peak detection algorithm rather than a dedicated one tailored for ECG data (\eg, Pan-Tompkins \cite{pan1985real}). The underlying reason is that IoT knowledge and programs only occupy a small proportion of the training dataset of code LLMs. As a result, despite being presented with abundant IoT terminologies in the prompt, LLMs still tend to prioritize and respond with more general words, due to their higher similarity (shorter distance in Fig.~\ref{fig:word_vector_space}) within the vector representation space.


LLM+RAG methods address this by retrieving domain knowledge for reference and establishing multiple cascaded agents to facilitate information transfer among modules. For example, in Fig.~\ref{fig:autoiot_limitation}, multiple LLM-based agents are employed for different tasks during development (\ie, domain knowledge retrieval, task planning, coding, and debugging). Sophisticated prompt design and meticulously structured intermediate outputs are necessary to ensure that one agent's output can be accurately parsed and interpreted by another agent. We conduct a preliminary experiment by prompting MapCoder \cite{islam2024mapcoder}, a multi-agent-based LLM+RAG framework, to synthesize programs for R-peak detection. We repeatedly generate 100 distinct versions of the programs and analyze them through code review and execution. Surprisingly, we find that only 28\% of the programs adopt appropriate IoT-related algorithms to perform R-peak detection. This is because LLM+RAG requires sophisticated RAG design and user prompts. Otherwise, the retrieved knowledge is less accurate and relevant to the IoT context, and LLMs may fail to focus on the IoT domain and still provide simple and general solutions \cite{wu2024faithful}. Moreover, this cascading process inevitably introduces noise and propagates errors \cite{wu2023next}, leading to a long self-recover time. More importantly, cloud LLMs suffer from bad network conditions, high costs, and privacy concerns.

To overcome these challenges, \textit{GPIoT} fine-tunes SLMs on IoT-specialized text-generation datasets, as SLMs have smaller sizes and can be locally deployed. Additionally, by steering the parameter distribution towards the IoT domain via tuning, SLMs can focus on IoT-related semantic context, generating highly relevant responses that follow pre-defined formats with enhanced stability.

\vspace{-5pt}
\subsection{Preliminary Experiments \& Findings}
\label{s:preliminary}

\vspace{-2pt}
We conduct some preliminary experiments by separately fine-tuning two SLMs on our manually constructed datasets (\S~\ref{sec:augmentation}), \ie, the task decomposition dataset (TDD) and the code generation dataset (CGD). TDD aims to enhance TDSLM's capability to break a \textit{problem statement} proposed by the user into multiple \textit{decomposed tasks} described in natural language. CGD aims to enhance CGSLM's ability to generate \textit{code \& documentation} for the user based on the \textit{decomposed tasks}. However, we find it extremely challenging to ensure the correctness of the generated code. Note that we use Llama2-13b \cite{touvron2023llama} as the default SLM for demonstration purpose.

\begin{figure}[t]
\vspace{-10pt}
\setlength{\abovecaptionskip}{-5pt}
\subfigtopskip=-4pt
\subfigcapskip=-4pt
    \centering
    \subfigure[BLEU scores]{
        \includegraphics[width=0.222\textwidth]{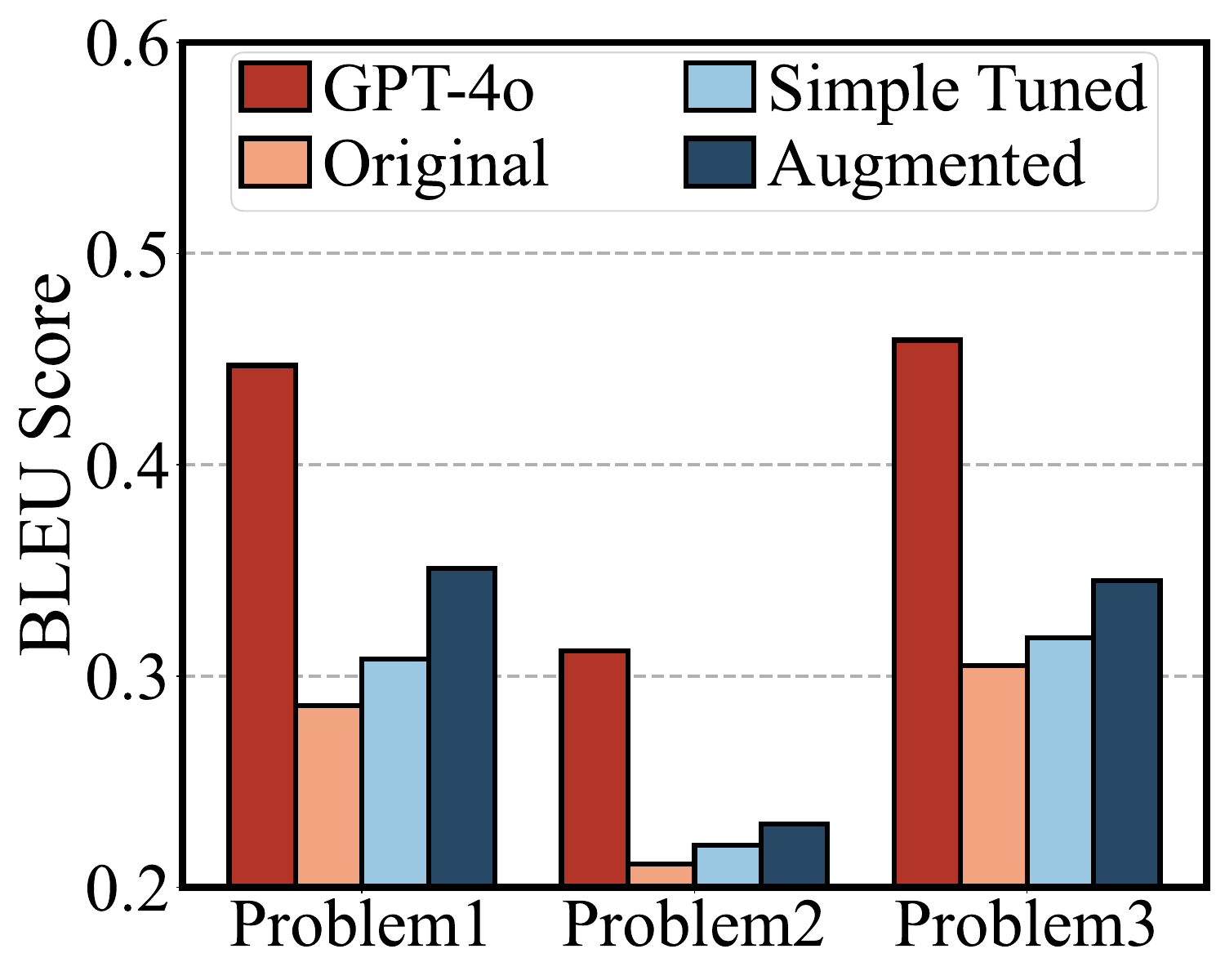}
        \label{fig:data_quality}
    }
    \centering
    \subfigure[Requirement-specification gap]{
        \label{fig:requirement_gap}
        \includegraphics[width=0.227\textwidth]{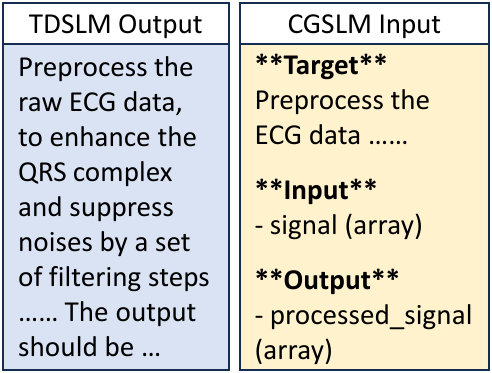}
    }
`    \caption{(a) Directly tuning SLM with simple augmentation only yields small improvements; (b) Gap between SLMs.}
    \vspace{-15pt}
\end{figure}

\noindent\textbf{Lack of high-quality data.}
Directly fine-tuning SLMs incurs little performance gain, even with data augmentation. 
We first deploy four models: GPT-4o, the original SLM, the SLM tuned on TDD, and the SLM tuned on augmented TDD via Evol-Instruct \cite{xu2023wizardlm}. Then, we randomly select three IoT problems from TDD and input them into the four models to obtain a set of responses.
Next, we measure the similarity between the generated responses and the human-crafted references as ground truth (\textit{decomposed tasks}) using the BLEU score \cite{papineni2002bleu}, where a larger value indicates higher semantic similarity. As shown in Fig.~\ref{fig:data_quality}, GPT-4o achieves the highest score with an acceptable value for such a text-generation task. However, the scores achieved by tuned SLMs increase slightly even with augmented TDD. Further analysis reveals that the solutions provided by the SLMs are either irrelevant to the IoT domain or contain hallucinations. This is because traditional augmentation methods (\eg, Evol-Instruct) focus on augmenting linguistic characteristics of the original text data, which may fall short of effectively capturing intricate relationships among IoT terminologies. \textit{This motivates us to design an IoT-tailored text augmentation method to enhance the quantity, quality, and diversity of the original dataset.}


\noindent\textbf{Domain misalignment.} 
Since TDSLM and CGSLM are tuned on distinct datasets for different tasks, domain misalignment occurs when used in tandem. Specifically, we feed the task descriptions generated by TDSLM into CGSLM to synthesize corresponding programs for each sub-task. Surprisingly, we find that only 53.4\% of the programs can be successfully executed without bugs and only 10.6\% of the programs adopt IoT-specialized algorithms for the IoT tasks. The main reason is that the two SLMs develop expertise in different domains with knowledge inconsistency during tuning, hindering the seamless integration of task decomposition and code generation. As a result, the responses generated by TDSLM may fall outside the scope that CGSLM can handle. \textit{This motivates us to develop a knowledge-sharing strategy between the two SLMs during tuning so that they can reach a consensus when handling IoT tasks.}

\noindent\textbf{Format incompatibility.} TDSLM's outputs (\textit{decomposed tasks}) are described in natural language while CGSLM's inputs (\textit{task specifications}) should be well structured (Fig.~\ref{fig:requirement_gap}).
When we directly feed TDSLM's output into CGSLM, only 23.6\% of the synthesized programs can be successfully executed. The rest exhibits higher uncertainty with a lack of confidence in mapping the input task specification to the desired code \cite{zhao2021calibrate}. The reason is that CGSLM is more sensitive to well-formatted inputs as it has been tuned on our dataset with structured text. Though directly tuning TDSLM to generate well-structured task specifications can be a solution, we find it challenging due to the limited language processing capabilities of SLMs, which cannot be sufficiently enhanced through tuning alone. \textit{This motivates us to develop a method to convert the task descriptions in natural language into well-organized specifications.}


To address the above challenges, we propose three key technical modules, \ie, an IoT-oriented text data augmentation method, a Parameter-Efficient Co-Tuning (PECT) paradigm with a multi-path LoRA pipeline, and a requirement transformation module. 

\begin{figure*}[t]
\vspace{-12pt}
\setlength{\abovecaptionskip}{2pt}
    \centering
    \includegraphics[width=\textwidth]{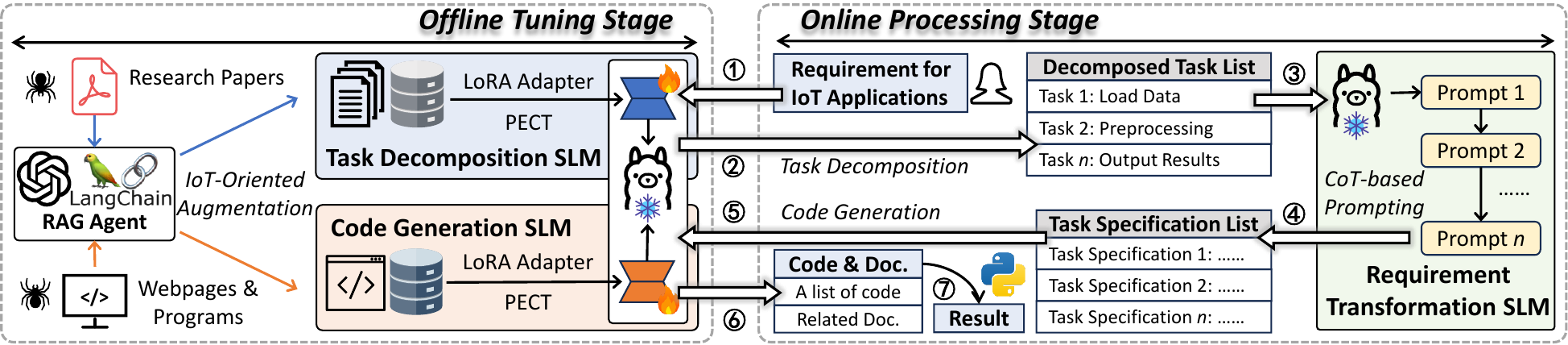}
    \caption{The system overview and workflow of \textit{GPIoT} (All the local SLMs share the same foundation model).}
    \label{fig:system_overview}
    \vspace{-15pt}
\end{figure*}

\section{System Overview}
\label{sec:overview}

\begin{figure}[t]
\vspace{4pt}
\setlength{\abovecaptionskip}{2pt}
    \centering
    \includegraphics[width=0.45\textwidth]{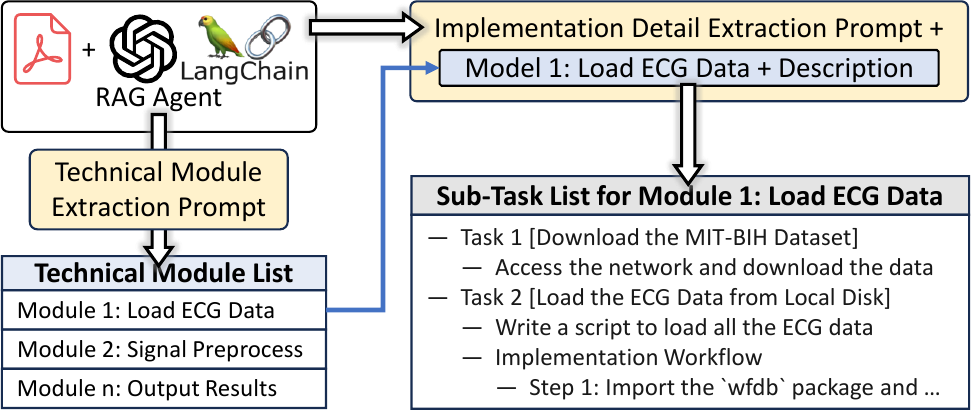}
    \caption{Task decomposition dataset construction.}
    \label{fig:task_data_formatting}
    \vspace{-15pt}
\end{figure}

Fig.~\ref{fig:system_overview} illustrates the overall architecture of \textit{GPIoT}, consisting of an offline tuning stage and an online processing stage.

\noindent\textbf{Offline Stage.} The offline tuning stage (the left part in Fig.~\ref{fig:system_overview}) constructs two IoT-specialized datasets and fine-tunes TDSLM and CGSLM, which will be used for task decomposition and code generation in the online stage, respectively. We first build a RAG agent to extract knowledge and programs from various IoT-related public sources (\eg, websites and articles) to \textbf{construct high-quality datasets}. 
Then, we augment the datasets by adopting our IoT-oriented augmentation method (\S~\ref{sec:augmentation}) to enhance their quantity, quality, and diversity. Note that the RAG agent is only used for high-quality dataset construction during the offline stage. 
With the two augmented datasets, we fine-tune two SLMs via our PECT paradigm, where certain model parameters are collaboratively tuned through a multiple-path LoRA pipeline with two projection layers for task decomposition and code generation, respectively. Our PECT paradigm \textbf{mitigates the domain misalignment} between TDSLM and CGSLM with facilitated knowledge transfer and sharing.


\noindent\textbf{Online Stage.} The online stage (the right part in Fig.~\ref{fig:system_overview}) aims to synthesize IoT-specific programs based on the user requirement for an IoT application development.
Specifically, \textit{GPIoT} first leverages \textit{Task Decomposition SLM} (TDSLM) to decompose the IoT application into multiple manageable sub-tasks with detailed descriptions (\ding{172}$\sim$\ding{173}). Next, through CoT-based prompting techniques, the sub-task descriptions will be gradually \textbf{transformed into well-structured specifications} by \textit{Requirement Transformation SLM} (RTSLM) (\ding{174}$\sim$\ding{175}). Next, for each sub-task, \textit{Code Generation SLM} (CGSLM) accordingly generates a code snippet with documentation (\ding{176}$\sim$\ding{177}). Users can execute the code sequentially to realize the IoT application based on the instructions from the documentation (\ding{178}). 

\noindent\textbf{SLM Considerations}. We consider SLMs as open-source models that can be locally deployed and operated efficiently on commodity GPUs (\eg, RTX 4070 Ti). This aligns with the practical constraints of normal users, where local models offer advantages in terms of cost, privacy, and independence from the cloud. Note that although there are three SLMs working simultaneously, \textit{they share the same foundation model and differ only in some additional tunable parameters}, which only occupy 1\% of all the parameters. Such a low-cost tuning and inference process stems from our PECT paradigm, avoiding significant overhead when deploying \textit{GPIoT} on local devices.

\vspace{-5pt}
\section{System Design}

\subsection{Data Collection \& Augmentation}
\label{sec:augmentation}
\vspace{-2pt}
Since there are no text-generation datasets in the IoT domain, we need to first construct a task decomposition dataset (TDD) and a code generation dataset (CGD). Note that our datasets contain Q\&A pairs in textual form, fundamentally differing from conventional IoT datasets that typically contain pairs of sensor data and labels.

\vspace{-3pt}
\subsubsection{\textbf{Task Decomposition Dataset.}}
\label{sec:TDD}
TDD contains pairs of "\textit{problem statement} $\rightarrow$ \textit{decomposed tasks}", aiming to enhance TDSLM's task decomposition ability for IoT problems.
The construction process consists of three stages: raw IoT-related text data collection, data formatting, and IoT-oriented text data augmentation.

\noindent\textbf{Raw Data Collection.} IoT-related research papers contain a huge quantity of high-quality SOTA applications and algorithms. Moreover, the systems proposed are comprehensive and functional, which can be decomposed into multiple modules with clear motivation and implementation details. Therefore, we download IoT-related papers from several public literature databases\footnote{We download papers from public databases via our institution's certification. The papers are for research only, adhering to ethical standards.} as our high-quality data sources, covering a wide range of IoT topics, such as communication, wireless sensing, edge computing, \etc.

\noindent\textbf{Data Formatting.}
We need to extract IoT knowledge from the papers and format it to pairs of "\textit{problem statement} $\rightarrow$ \textit{decomposed tasks}".
Intuitively, we can regard the proposed system in each paper as an IoT problem, with its technical modules as the corresponding decomposed tasks.
However, two challenges occur if we directly use such "\textit{System} $\rightarrow$ \textit{technical modules}" pairs for tuning: 
1) These module descriptions are typically lengthy, which exceed the context length of SLMs \cite{chen2023extending}. 2) These modules are still sophisticated, often containing multiple sub-systems. TDSLM may struggle to extract IoT-specialized technical concepts and accurately generate manageable components.
To tackle this, our insight is that we regard each module as an individual problem, which can be further decomposed into several manageable sub-tasks. The disintegrated sub-tasks can be easily handled by TDSLM with reduced context length.

\begin{figure}[t]
\vspace{-8pt}
\setlength{\abovecaptionskip}{-4pt}
\subfigtopskip=-4pt
\subfigcapskip=-4pt
    \centering
    \subfigure[Technical module extraction]{
        \label{fig:technical_module_prompt}
        \includegraphics[width=0.225\textwidth]{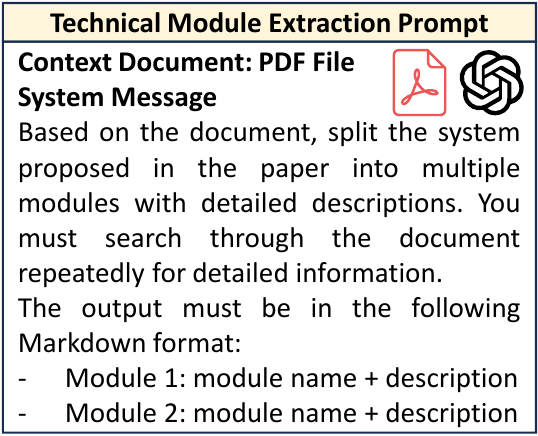}
    }
    \centering
    \subfigure[Implementation detail extraction]{
        \label{fig:implementation_detail_prompt}
        \includegraphics[width=0.225\textwidth]{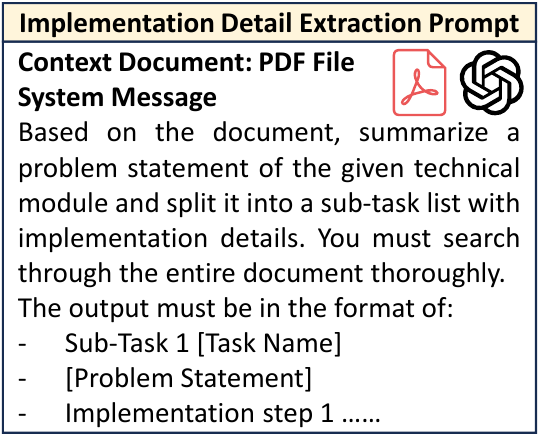}
    }
    \caption{Prompts for paper information extraction.}
    \vspace{-10pt}
\end{figure}

\begin{figure}
\setlength{\abovecaptionskip}{2pt}
    \centering
    \includegraphics[width=0.42\textwidth]{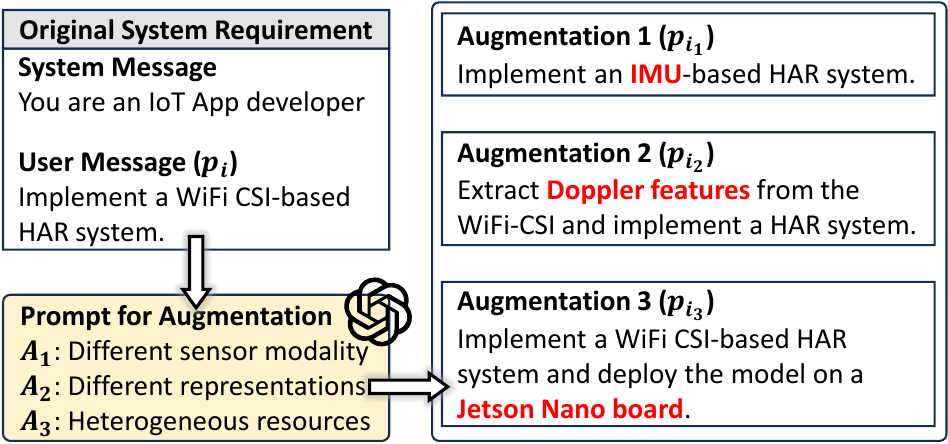}
    \caption{Examples of IoT-oriented data augmentation with different modalities, representations, and resource budgets}.
    \label{fig:augmentation_example}
    \vspace{-18pt}
\end{figure}

\begin{figure}[t]
\vspace{-8pt}
\setlength{\abovecaptionskip}{-5pt}
\subfigtopskip=-4pt
\subfigcapskip=-4pt
    \centering
    \subfigure[Task decomposition data sample]{
        \includegraphics[width=0.225\textwidth]{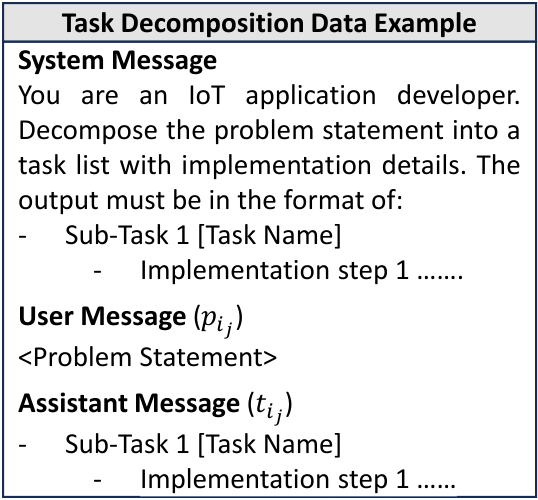}
        \label{fig:task_decomposition_data_sample}
    }
    \centering
    \subfigure[Augment sensor modality]{
        \includegraphics[width=0.225\textwidth]{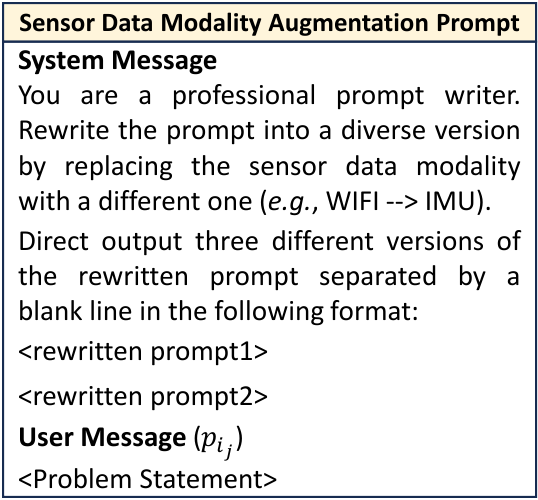}
        \label{fig:augmentation_prompt}
    }
    \caption{(a) Tuning data sample for task decomposition. (b) Prompt for sensor modality augmentation.}
\vspace{-12pt}
\end{figure}

Fig.~\ref{fig:task_data_formatting} shows the entire process of how we extract pairs of "\textit{problem statement} $\rightarrow$ \textit{decomposed tasks}" from the papers.
Specifically, we first build a RAG agent by combining the downloaded papers with an LLM (GPT-4o). Based on the provided context documents, we then prompt (Fig.~\ref{fig:technical_module_prompt}) the agent to split the proposed system in the paper into multiple technical modules with detailed descriptions. Next, for each technical module, we prompt (Fig.~\ref{fig:implementation_detail_prompt}) the agent to further decompose it into several sub-tasks with detailed implementation steps.
As such, we encapsulate the \textit{problem statement} $p_i$ of each technical module and the corresponding sub-tasks $t_i$ into a Q\&A pair $Q_i$ to construct a raw dataset $\mathcal{D}_t$:
\begin{equation}
\small
\setlength{\abovedisplayskip}{0pt}
\setlength{\belowdisplayskip}{0pt}
    \mathcal{D}_t = \{Q_1, Q_2, \cdots, Q_{n_t}\}, \ Q_i = (p_i, t_i)
\end{equation}
where $n_t$ is the total number of technical modules from all the papers. Fig.~\ref{fig:task_decomposition_data_sample} shows a data sample from the dataset. Note that each sub-task is separated by a blank line, allowing us to parse and split $t_i$ into multiple task description strings for further code generation in a divide-and-conquer way.


\noindent\textbf{IoT-Oriented Data Augmentation.}
As revealed in \S~\ref{s:preliminary}, existing text augmentation methods are ineffective in the IoT domain as they focus on expanding language characteristics rather than IoT knowledge.
As a result, IoT terminologies are still assigned lower priority during inference, preventing the tuned model from generating IoT-specialized solutions.
To address this, we propose a novel IoT-oriented data augmentation method that considers unique properties of IoT applications, \ie, sensor modality, data representation, and system resource heterogeneity, as shown in Fig.~\ref{fig:augmentation_example}.

Our augmentation considers three aspects: 1) \textit{Sensor modality}. For the same IoT problem, we can use different sensor modalities. For instance, to implement human activity recognition (HAR), we can utilize IMU data, WiFi CSI, \etc. 2) \textit{Data representation}. For the same modality, we can leverage distinct data representations to achieve the same task. For example, we can use WiFi CSI, 2D spectrograms, or extracted Doppler features to implement HAR. 3) \textit{Resource heterogeneity}. For the same task, various IoT devices require heterogeneous system resources. When deploying an AI model, smartphones typically have less memory than PCs, requiring model optimization methods. Based on the three aspects, we prompt (Fig.~\ref{fig:augmentation_prompt}) GPT-4o to rewrite and augment each \textit{problem statement} from $\mathcal{D}_t$. To generate reference decomposed tasks, we build a search agent to retrieve relevant IoT domain knowledge and prompt it to produce results. We then manually filter out incorrect results and craft the formats (\ie, each sub-task is separated by a blank line as aforementioned). The augmented dataset $\mathcal{D}_t^\prime$ is:
\begin{equation}
\small
\setlength{\abovedisplayskip}{0pt}
\setlength{\belowdisplayskip}{0pt}
    \mathcal{D}_t^\prime = \bigcup_j^3{\{(p_{i_j}, t_{i_j})\ \vert\ p_{i_j}=A_j(p_i),\ t_{i_j}=G(p_{i_j})\}},\ \forall p_i\in \mathcal{D}_t
\end{equation}
where $A_j(\cdot)$ is the $j$-th type of augmentation operation and $G(\cdot)$ is the black-box function of GPT-4o.


\noindent\textbf{Remark.} We take the diversity of both language expression and IoT characteristics into account, demonstrating significant performance improvement in task decomposing (\S~\ref{s:ablation}). Note that the data collection and augmentation processes are both performed offline.

\begin{figure}
\setlength{\abovecaptionskip}{2pt}
    \centering
    \includegraphics[width=0.45\textwidth]{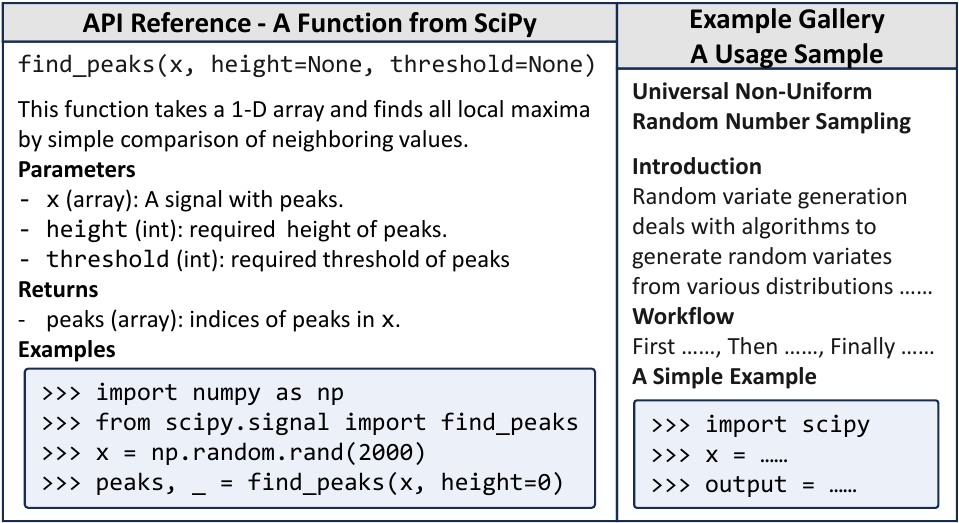}
    \caption{An example of a Python package's website.}
    \label{fig:package_information_example}
    \vspace{-18pt}
\end{figure}

\vspace{-3pt}
\subsubsection{\textbf{Code Generation Dataset.}}
CGD contains pairs of "\textit{task specification} $\rightarrow$ \textit{code \& documentation}", aiming to enhance CGSLM's ability in generating IoT-related code for decomposed tasks.
The data construction includes two stages: raw data collection and target diversity-aware augmentation for different code generation tasks.

\noindent\textbf{Raw Data Collection.} 
Open-source IoT-related Python\footnote{We focus on Python since it is a cross-platform programming language.} packages (\eg, SciPy \cite{2020SciPy-NMeth}) contain abundant hand-crafted IoT algorithms and applications with high performance, which can serve as our data sources. Thus, we first collect numerous public repositories from GitHub and extract Python packages they used, covering areas of signal processing, machine learning, and data processing (IoT data I/O and visualization). We then build a web crawler to automatically retrieve information from each package's official website.

\begin{figure*}[t]
\vspace{-8pt}
\setlength{\abovecaptionskip}{-5pt}
\subfigtopskip=-4pt
\subfigcapskip=-4pt
    \centering
    \subfigure[Module description]{
        \includegraphics[width=0.227\textwidth]{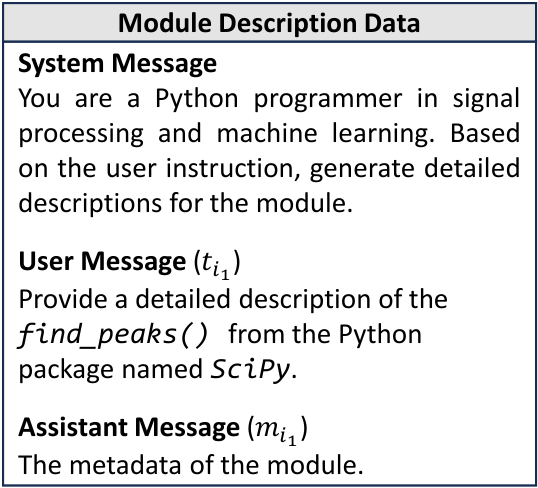}
        \label{fig:module_description}
    }
    \centering
    \subfigure[Module implementation]{
        \includegraphics[width=0.227\textwidth]{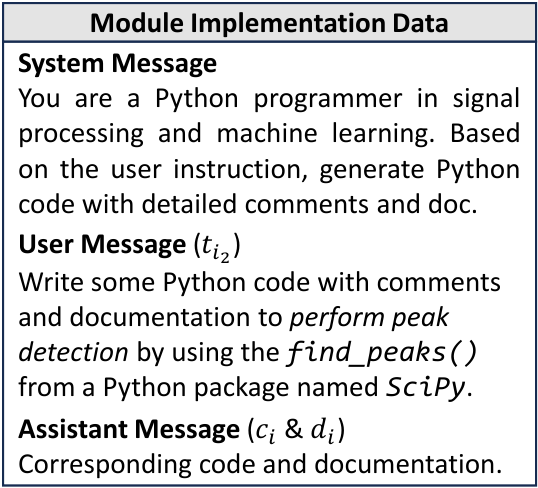}
        \label{fig:module_implementation}
    }
    \centering
    \subfigure[Task specification]{
        \includegraphics[width=0.227\textwidth]{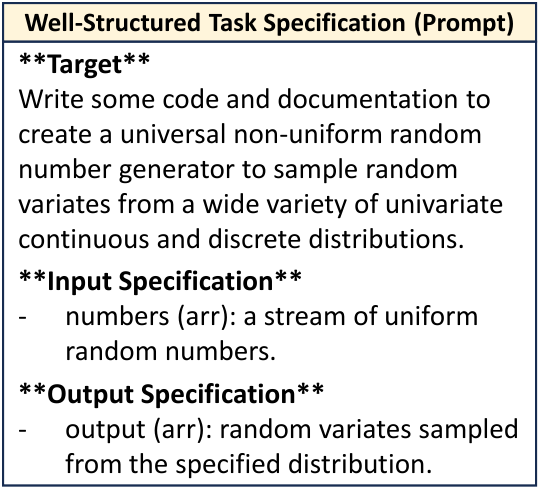}
        \label{fig:task_specification_sample}
    }
    \centering
    \subfigure[Traditional LoRA tuning]{
        \includegraphics[width=0.22\textwidth]{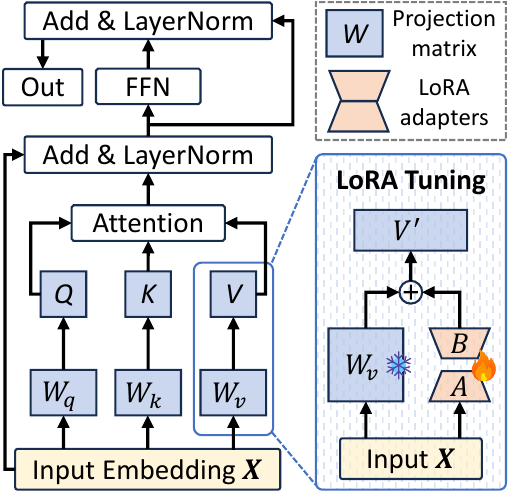}
        \label{fig:traditional_lora_tuning}
    }
    \caption{(a) \& (b) Two tuning data samples. (c) Well-structured task specification. d) Traditional LoRA tuning.}
    \vspace{-18pt}
\end{figure*}

A package's website (Fig.~\ref{fig:package_information_example}) typically contains two parts: 1) \textit{API reference} includes a list of modules (\ie, functions and classes) with comprehensive guidance on how to use them effectively in code. For example, \verb|find_peaks()| is a function from the SciPy package, which identifies the local maxima (peaks) in an input signal array and returns the indices of the peaks. We denote the detailed information of each module as its metadata, $m_i$. 2) \textit{Example gallery} provides practical usage samples of how to use various modules and features of the package to implement specific algorithms. For instance, a usage sample provides detailed documentation with code about performing universal non-uniform random number sampling using the SciPy package in an end-to-end manner. We denote the detailed information of each usage sample as its metadata, $u_j$. By combing these two types of metadata, we form a raw dataset $\mathcal{D}_c$:
\begin{equation}
\setlength{\abovedisplayskip}{0pt}
\setlength{\belowdisplayskip}{0pt}
    \small
    \mathcal{D}_c = \{m_i\ \vert\ \forall i\in\{1, 2, \cdots, n_m\}\} \cup \{u_j\ \vert\ \forall j\in\{1, 2, \cdots, n_u\}\}
\end{equation}
where $n_m$ and $n_u$ is the total number of modules and usage samples.

\noindent\textbf{Target Diversity-Aware Augmentation.}
This augmentation aims to enhance the diversity of the metadata in $\mathcal{D}_c$. We target each module in the packages for two text-generation tasks: 
1) \textit{Module Description}: providing detailed descriptions of a module and 2) \textit{Module Implementation}: writing code \& documentation to demonstrate usage samples of the module.
aSince the example gallery already contains abundant algorithms with sample code and detailed descriptions, they can be directly used as code generation tasks.

\noindent1) \textit{Module Description.} We format the task specification to "Provide detailed descriptions of \verb|<module>| from \verb|<package>|." The corresponding reply contains the module metadata in a pre-defined format. This Q\&A mapping relation from a \textit{task specification} $t_i$ to a \textit{module description} $m_i$ is expressed as:
\begin{equation}
\small
\setlength{\abovedisplayskip}{0pt}
\setlength{\belowdisplayskip}{0pt}
    \mathcal{D}_1 = \{t_i\rightarrow m_i\}
\end{equation}
Such a "\textit{task specification} $\rightarrow$ \textit{module description}" mapping can teach CGSLM to be familiar with the module's information, strengthening the semantic correlation between the module's name and the detailed descriptions. Fig.~\ref{fig:module_description} illustrates a data sample from $\mathcal{D}_1$.

\noindent2) \textit{ Module Implementation.} We format the task specification to "Write some Python code with comments and documentation to perform \verb|<target>| by using \verb|<module>| from \verb|<package>|." The corresponding reply contains the sample code and documentation that provides the workflow and guidance on how to execute the code. To obtain well-structured documentation, we prompt GPT-4o to format the module's metadata into Markdown. This Q\&A mapping relation from a \textit{task specification} $t_i$ to the corresponding \textit{module implementation} (\ie, code $c_i$ and documentation $d_i$) is:
\begin{equation}
\setlength{\abovedisplayskip}{0pt}
\setlength{\belowdisplayskip}{0pt}
    \small
    \mathcal{D}_2 = \{t_i\rightarrow (c_i, d_i)\ \vert\ d_i=G(m_i)\}
\end{equation}
Such a "\textit{task specification} $\rightarrow$ \textit{module implementation}" mapping relationship aims to enhance CGSLM's capability in generating code and documentation according to the module specification. Fig.~\ref{fig:module_implementation} shows a data sample from $\mathcal{D}_2$.

\noindent3) \textit{Example Implementation.} We format the task specification to a well-structured Markdown format as shown in Fig.~\ref{fig:task_specification_sample}, including the task target and the I/O specifications for the expected code. Correspondingly, we prompt GPT-4o to convert the usage sample's metadata into well-structured documentation. This Q\&A mapping relation from a \textit{task specification} $t_j$ to the code $c_j$ and documentation $d_j$ can be expressed as:
\begin{equation}
\setlength{\abovedisplayskip}{0pt}
\setlength{\belowdisplayskip}{0pt}
    \small
    \mathcal{D}_3 = \{t_j\rightarrow (c_j, d_j)\ \vert\ d_j=G(u_j)\}
\end{equation}
This aims to enhance CGSLM's ability in generating IoT-related code and detailed documentation following well-structured task specifications. Ultimately, by concatenating all three augmented datasets, the final CGD $\mathcal{D}_c^\prime$ becomes:
\begin{equation}
\setlength{\abovedisplayskip}{0pt}
\setlength{\belowdisplayskip}{0pt}
\small
    \mathcal{D}_c^\prime = \mathcal{D}_1\{t_i\rightarrow m_{i_1}\}\cup\mathcal{D}_2\{t_i\rightarrow (c_i, d_i)\}\cup\mathcal{D}_3\{t_j\rightarrow (c_j, d_j)\}
\end{equation}

\vspace{-3pt}
\subsubsection{\textbf{\textit{IoTBench}}}
\label{sec:iotbench}
To evaluate LLMs' abilities in task decomposition and code generation for IoT applications, we create \textit{IoTBench}, a benchmark of text-generation tasks in the IoT domain. Specifically, we choose 100 samples from TDD and CGD with manually created test cases, covering various IoT topics (\eg, signal processing, edge AI, \etc.). All the selected data samples are first manually filtered to ensure correctness and relevance to the IoT domain. Then, we format the sub-tasks separated by a blank line in between.
Note that although many SOTA benchmarks (\eg, HumanEval \cite{chen2021evaluating}) can also evaluate LLMs' code generation abilities, they are not tailored to IoT tasks. Besides, the data in \textit{IoTBench} is excluded from the tuning processes (\S~\ref{sec:tuning}) to test the generalizability of the tuned SLMs.

\vspace{-5pt}
\subsection{Parameter-Efficient Co-Tuning (PECT)}
\label{sec:tuning}

With the two augmented datasets ($\mathcal{D}_t^\prime$ and $\mathcal{D}_c^\prime$), our next step is to fine-tune TDSLM and CGSLM to enhance their ability in task decomposition and code generation, respectively. In the following, we first introduce the traditional LoRA tuning method \cite{hu2021lora} for SLMs, and then explain our Parameter-Efficient Co-Tuning paradigm.

\vspace{-3pt}
\subsubsection{\textbf{LoRA Tuning.}}
Fig.~\ref{fig:traditional_lora_tuning} shows the Low-Rank Adaptation (LoRA) tuning process of a Transformer block \cite{vaswani2017attention}. Specifically, each Transformer block in an LLM contains two main components: a self-attention mechanism and a feed-forward network (FFN), both of which are followed by residual connections and layer normalization. The self-attention features three tunable weight matrices ($W_q$, $W_k$, and $W_v$) to capture contextual relationships between input embeddings, while the FFN processes the outputs from the attention mechanism to refine the feature representations. In conventional LLM full-tuning, the entire weight matrices are updated, leading to extensive GPU memory requirements and high computational costs. Instead of fully updating the weight matrices, LoRA reduces the number of tunable parameters \cite{dettmers2024qlora, ding2023sparse}, where two low-rank matrices $\boldsymbol{A}$ and $\boldsymbol{B}$ (\ie, LoRA adapters) are inserted alongside the weight matrix.
Given an input $\boldsymbol{X}$, the tuning process can be expressed as:
\begin{equation}
\setlength{\abovedisplayskip}{0pt}
\setlength{\belowdisplayskip}{0pt}
\small
    \boldsymbol{V}^\prime = (\boldsymbol{W}_v + \boldsymbol{BA})\cdot \boldsymbol{X}
\end{equation}
This reduces the computational burden by only updating the smaller low-rank matrices $\boldsymbol{A}$ and $\boldsymbol{B}$, significantly cutting down resources.


However, as demonstrated in \S~\ref{s:preliminary}, domain misalignment arises when separately tuning TDSLM and CGSLM using this vanilla LoRa method. This is because they focus on two distinct text-generation tasks with different semantic attention, thereby hindering \textit{GPIoT} from synthesizing IoT-related programs. To tackle this issue, we propose a parameter-efficient co-tuning (PECT) paradigm. \textbf{\textit{Unlike conventional LoRA tuning that tunes adapters separately, PECT enables collaborative fine-tuning of several SLMs with a shared base model but with different LoRA adapters.}} PECT features a Multi-Path LoRA Pipeline (MPLP) and two lightweight projection layers, which can promote information sharing between TDSLM and CGSLM, thereby narrowing the semantic comprehension gap between task decomposition and code generation.


\begin{figure}[t]
\vspace{-12pt}
\setlength{\abovecaptionskip}{2pt}
    \centering
    \includegraphics[width=0.45\textwidth]{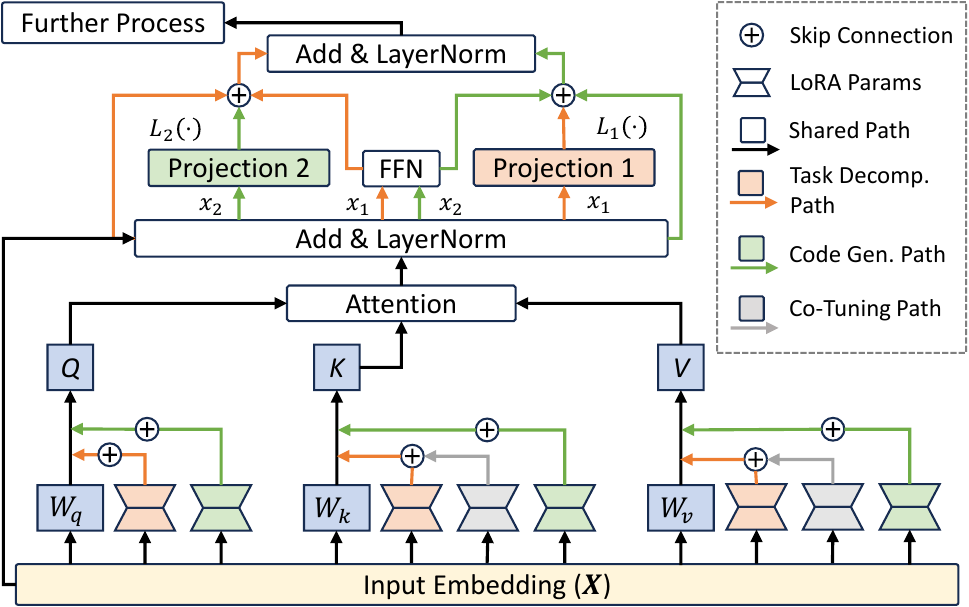}
    \caption{PECT in one Transformer block with both independently and collaboratively tuned LoRA adapters.}
    \label{fig:multipath_LoRA}
\vspace{-18pt}
\end{figure}

\vspace{-3pt}
\subsubsection{\textbf{Multi-Path LoRA Pipeline.}}
 MPLP selects a subset of shared LoRA adapters to be collaboratively tuned by TDSLM and CGSLM, with another adapter set independently tuned.




\noindent\textbf{Pipeline Construction.} In the lower part of Fig.~\ref{fig:multipath_LoRA}, we create three pipelines of LoRA adapters in each Transformer block. Two pipelines (the orange one and the green one) are independently tuned by TDSLM and CGSLM with respect to TDD and CGD. The other pipeline (the gray one) is co-tuned on both TDD and CGD. For example, given a data sample from TDD, only the orange LoRA adapters and gray LoRA adapters are updated, as shown in Fig.~\ref{fig:multipath_LoRA}. Note that we only assign the shared adapters beside the key and value weight matrices ($\boldsymbol{W}_k$, $\boldsymbol{W}_v$). 
The insight behind this is that the value vector provides the information to be activated based on the key vector \cite{vaswani2017attention}. In other words, the mapping from \textit{problem statement} to \textit{decomposed tasks} and the mapping from \textit{task specification} to \textit{code \& documentation} are determined by the key and value vectors in TDSLM and CGSLM, respectively. Domain misalignment is thus caused by such different mapping relations during tuning on distinct datasets with disparate semantic focuses. Therefore, by co-tuning the shared adapters integrated into the key and value vectors, the mapping relations will be shared between the two SLMs, allowing TDSLM's outputs to align with CGSLM's scope.

\noindent\textbf{Co-Tuning.} In Fig.~\ref{fig:multipath_LoRA}, we designate the orange line as the task decomposition path (TDP), through which only data from TDD will pass. The green line is the code generation path (CGP), through which only data from CGD will pass. The grey line represents the co-tuning path, through which all data will pass. During co-tuning, the LoRA adapters will be tuned either independently or collaboratively, depending on the path they occupy. Specifically, take the adapter alongside the projection matrix $\boldsymbol{W}_k$ as an example, the key vectors after projection in the two paths are calculated by:
\begin{equation}
\small
\setlength{\abovedisplayskip}{0pt}
\setlength{\belowdisplayskip}{0pt}
\label{eq:multi_path_lora}
\begin{array}{l}
	\boldsymbol{K}_1 = (\boldsymbol{W}_k+\boldsymbol{B}_1\boldsymbol{A}_1 + \lambda\cdot\boldsymbol{B}_c\boldsymbol{A}_c)\cdot \boldsymbol{X} \\
	\boldsymbol{K}_2 = (\boldsymbol{W}_k+\boldsymbol{B}_2\boldsymbol{A}_2 + (1-\lambda)\cdot\boldsymbol{B}_c\boldsymbol{A}_c)\cdot \boldsymbol{X}
\end{array}
\end{equation}
where $\boldsymbol{X}$ is the input text embedding, $\boldsymbol{K}_1$ and $\boldsymbol{K}_2$ are the key vectors within TDP and CGP, respectively. $\boldsymbol{B}_1\boldsymbol{A}_1$ and $\boldsymbol{B}_2\boldsymbol{A}_2$ are the parameters of LoRA adapters independently tuned within the two paths, respectively. $\boldsymbol{B}_c\boldsymbol{A}_c$ are the LoRA adapters collaboratively tuned by the two paths. $\lambda$ is a hyper-parameter to balance the data flow between the two paths. During the co-tuning process, we first randomly sample data from TDD and CGD. Next, if the data is sampled from TDD, it will pass through TDP; otherwise, it will pass through CGP. We then calculate the loss and update the corresponding LoRA adapters based on the source of the data sample.

\noindent\textbf{Remarks.} By orchestrating the independent and collaborative tuning paths, MPLP dismantles the information barrier between TDSLM and CGSLM, fostering their consensus during inference and thereby alleviating the misalignment issue.

\begin{figure}[t]
\vspace{-8pt}
\setlength{\abovecaptionskip}{-5pt}
\subfigtopskip=-4pt
\subfigcapskip=-4pt
    \centering
    \subfigure[User target extraction]{
        \includegraphics[width=0.225\textwidth]{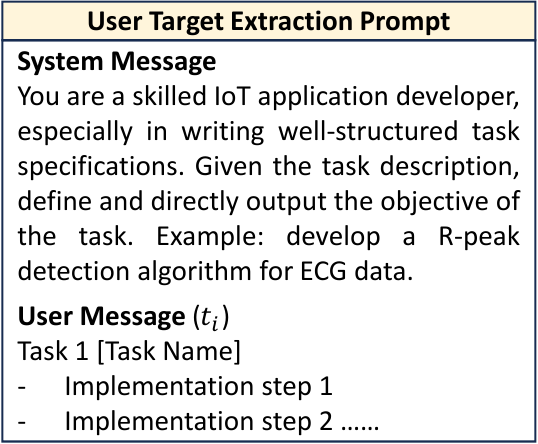}
        \label{fig:user_target_extraction}
    }
    \centering
    \subfigure[I/O specification extraction]{
        \includegraphics[width=0.225\textwidth]{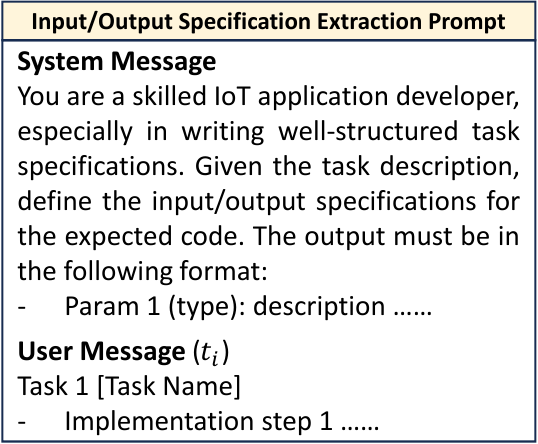}
        \label{fig:input_output_specification_extraction}
    }
    \caption{CoT-based prompts for RTSLM.}
\vspace{-18pt}
\end{figure}

\vspace{-3pt}
\subsubsection{\textbf{Projection Layers.}}
To further enhance knowledge sharing between the two SLMs during tuning, we create two projection layers for the two paths. We place the projection layers in parallel with the FFN layers in the Transformer block. As such, they can serve as extra FFNs that apply non-linear transformations to the attention representations, thereby enhancing token-level feature extraction and increasing the complexity of the model's learning capabilities. By receiving representations from the other path with enhanced non-linearity, the cross-domain IoT knowledge comprehension capabilities of the two SLMs will be further strengthened. Specifically, take TDP as an example, as shown in the upper part of Fig.~\ref{fig:multipath_LoRA}, with the obtained value (denoted as $x_1$) from the LayerNorm in TDP, we feed it into a projection layer $L_1(\cdot)$. The output is then added with the FFN's output $F(x_2)$ and $x_2$ in CGP. The sum will be sent to the next Transformer block for further processing. Such a knowledge transfer process can be expressed as:
\begin{equation}
\label{eq:projection_layer}
\small
\setlength{\abovedisplayskip}{0pt}
\setlength{\belowdisplayskip}{0pt}
    \begin{array}{l}
         x^\prime_1 = x_1 + F(x_1) + \gamma\cdot L_2(x_2)\\
         x^\prime_2 = x_2 + F(x_2) + (1-\gamma)\cdot L_1(x_1)
    \end{array}
\end{equation}
where $x^\prime_1$ and $x^\prime_2$ are the final output of the two paths. $x_1$ and $x_2$ are the input attention representations from TDP and CGP, respectively. $F(\cdot)$ represents the FFN layer, $L_1(\cdot)$ and $L_2(\cdot)$ are the projection layers in the two paths, and $\gamma$ is a hyper-parameter to balance the knowledge-sharing between the two paths. Note that each projection layer has the same architecture as the FFN, consisting of two fully connected layers and a non-linear SwiGLU function \cite{shazeer2020glu}.

\noindent\textbf{Remarks.} By combining independent and collaborative tuning of LoRA adapters with the projection layers, PECT optimizes the task-specific performance of TDSLM and CGSLM while minimizing domain conflicts. As a result, the decomposed tasks generated by TDSLM will have closer semantic alignment with CGSLM and thus can be better handled. Note that during both tuning and inference, the SLMs share the same foundation model architecture, with the only difference being the LoRA parameters shown in Fig.~\ref{fig:multipath_LoRA}. Therefore, our proposed PECT paradigm fundamentally differs from those traditional PEFT approaches that tune SLMs separately.

\vspace{-8pt}
\subsection{Requirement Transformation}
\label{sec:requirement_transformation}
\vspace{-2pt}
When cascading TDSLM and CGSLM together, a huge gap exists between TDSLM's outputs (\textit{decomposed task descriptions}) and CGSLM's inputs (\textit{task specifications}). The task descriptions are typically natural language while the task specifications are well-structured. Directly feeding the task descriptions into CGSLM will lead to sub-optimal performance of the generated code. To fill this gap, we leverage RTSLM to transform the descriptions into well-structured specifications. Considering that RTSLM has limited IoT knowledge comprehension ability, we enhance it with RAG and several CoT-based prompts to perform requirement transformation.

\noindent\textbf{RAG Construction.} To enhance RTSLM's ability in understanding IoT domain knowledge during requirement transformation, we first transform all the downloaded papers into a text embedding database. Then, armed with such an IoT knowledge database, we build a RAG agent based on RTSLM to retrieve relevant context for reference. As such, RTSLM can better comprehend and handle IoT terminologies in the task descriptions during requirement transformation.

\noindent\textbf{CoT Prompting.} Fig.~\ref{fig:task_specification_sample} shows an example of a well-structured task specification for code generation, consisting of three parts: task target, input and output specifications of the expected code. For each decomposed task $t_i$ generated by TDSLM, we prompt the agent to generate such well-structured specifications step-by-step. Specifically, we first prompt (Fig.~\ref{fig:user_target_extraction}) the agent to summarize a target for the task. Next, we further instruct (Fig.~\ref{fig:input_output_specification_extraction}) the agent to generate a list of parameter descriptions for the input and output of the expected code. Each single parameter description item contains the parameter name, the parameter type, and a brief explanation of its meaning. For example, "signal (numpy.ndarray): the raw ECG data collected from patients with noises." Finally, RTSLM reorganizes and formats the above information into a well-structured task specification, which will be further handled by CGSLM to generate corresponding code snippets and documentation.

\noindent\textbf{Remarks.} Note that tuning is excluded in this process since it only needs basic language comprehension and processing capabilities of RTSLM. Therefore, RTSLM shares the same base model without additional tunable LoRA parameters to perform the transformation.

\vspace{-5pt}
\section{Experiment Setup}

\vspace{-3pt}
\subsection{Implementation}
\vspace{-3pt}
\textbf{System Configurations.}
We deploy \textit{GPIoT} on an edge PC equipped with an RTX 4090 GPU (24 GB). We use selenium \cite{selenium} to create a web crawler for data retrieving from public websites. To perform data formatting and augmentation, we construct an agent based on GPT-4o and LangChain \cite{Chase_LangChain_2022}. For SLM tuning, we use a high-performance cloud server with an NVIDIA A100 GPU (80 GB). 

\noindent\textbf{Hyper-parameters.} TDD contains 36,098 pairs of "\textit{problem statement} $\rightarrow$ \textit{decomposed tasks}". CGD contains 35,419 pairs of "\textit{task specification} $\rightarrow$ \textit{code \& documentation}". Llama2-13b \cite{touvron2023llama} with INT8 quantization serves as the foundation model and is fine-tuned via LoRA \cite{hu2021lora}, with a rank of 64 and a dropout rate of 0.1. The number of tuning epochs is 5, with an initial learning rate of 0.0001, varied by a cosine learning rate scheduler. The $\lambda$ in Eq.~\ref{eq:multi_path_lora} and the $\gamma$ in Eq.~\ref{eq:projection_layer} are both set to 0.5 by default. The tuning process takes around 80 GPU hours. Since TDSLM, RTSLM, and CGSLM share the same foundation model, only about 16 GB of GPU memory is needed for the whole system, which is affordable for a commodity GPU \cite{pudipeddi2020training}.

\vspace{-8pt}
\subsection{IoT Applications}
\vspace{-2pt}
Considering the different technologies required during development, we select three IoT applications, focusing on healthcare and edge computing. 1) \textbf{Heartbeat Detection (HD)} is essential for continuously monitoring patient vitals with enhanced healthcare and ensuring timely intervention in case of abnormalities \cite{ouyang2024admarker}. We instruct \textit{GPIoT} to develop a heartbeat (R-peak) detection algorithm and test it on the MIT-BIH dataset \cite{moody2001impact}. 2) \textbf{Human Activity Recognition (HAR)} \cite{ji2024hargpt, ji2022sifall, cao2024finger, ji2023construct, adhikari2024misleep} deployed on edge devices is important for real-time analysis of daily human activities. We instruct \textit{GPIoT} to develop a WiFi-based HAR model using the WiAR dataset \cite{guo2019wiar} and deploy it on a Jetson Nano board that has limited resources \cite{ling2021rt}. 3) \textbf{Multimodal HAR} leverages different sensors to capture complementary information, thereby enhancing HAR systems' robustness and versatility \cite{fan2025diffusion}. We instruct \textit{GPIoT} to construct a multimodal HAR model based on the Harmony dataset \cite{ouyang2023harmony}, which contains three sensor modalities: audio \cite{wang2023audioguard}, depth camera, and radar \cite{cui2024talk2radar}.

\noindent\textbf{Notes:} HD requires signal processing methods, HAR demands technologies in both signal processing and machine learning, and multimodal HAR necessitates advanced multimodal processing algorithms. Though we use HD as an example for demonstration throughout the paper, all the tasks are unseen to \textit{GPIoT}.

\vspace{-5pt}
\section{Evaluation}
\vspace{-2pt}
\subsection{Metrics}
\label{sec:overall_metric}
\vspace{-2pt}
We compare the programs synthesized by \textit{GPIoT} and several baselines by measuring the following evaluation metrics.

\noindent\textbf{HD}. 1) \textit{Precision}: The fraction of correctly detected R-peaks out of all detected peaks: $\frac{TP}{TP+FP}$. 2) \textit{Recall rate}: The proportion of correctly detected R-peaks out of all actual R-peaks: $\frac{TP}{TP+FN}$. The larger these two metrics are, the more accurate the heartbeat detection becomes.

\noindent\textbf{HAR}. 1) \textit{Classification accuracy}: The portion of the test data that is correctly classified based on the label. A higher accuracy implies a more robust and accurate HAR model. 2) \textit{GPU memory usage}: The amount of GPU memory used during model inference. 3) \textit{Inference time}: The time it takes from feeding the data into the code to the generation of the recognition result. The less memory and inference time consumed, the more resource-efficient the HAR model is.

\vspace{-8pt}
\subsection{Baselines}
\label{sec:baseline}
\vspace{-2pt}
Given the same user problem, we input it into the following baselines to compare their performance with \textbf{\textit{GPIoT} (GT)}. 1) \textbf{GPT-4o (G4)} \cite{achiam2023gpt} is an advanced LLM from OpenAI, optimized for instruction following and code generation tasks. 2) \textbf{DeepSeek-Coder (DC)} \cite{guo2024deepseek} is a high-performance code LLM, particularly effective in understanding and generating programming code across various domains. 3) \textbf{CodeLlama-34b (CL)} \cite{roziere2023code} is a specialized version of the Llama designed to generate, understand, and assist with coding. 4) \textbf{WizardCoder-33b (WC)} \cite{luo2023wizardcoder} incorporates complex instruction fine-tuning by adopting evolving instructions. 5) \textbf{CodeQwen-7b (CQ)} \cite{bai2023qwen} is the Code-Specific version of Qwen1.5, which is a decoder-only LLM pre-trained on a large amount of data of programs. 6) \textbf{GitHub Copilot (GC)} \cite{github} is an AI-powered code generation tool that assists developers by suggesting code snippets and functions. 7) \textbf{MapCoder (MC)} \cite{islam2024mapcoder} is an LLM+RAG-based code generation framework that cascades multiple LLM-based agents to solve competitive problems, where GPT-4o is selected as the built-in LLM.

\begin{figure}
\vspace{-8pt}
\setlength{\abovecaptionskip}{-3pt}
\subfigtopskip=-3pt
\subfigcapskip=-3pt
    \centering
    \subfigure[Problem statement for HD]{
        \includegraphics[width=0.225\textwidth]{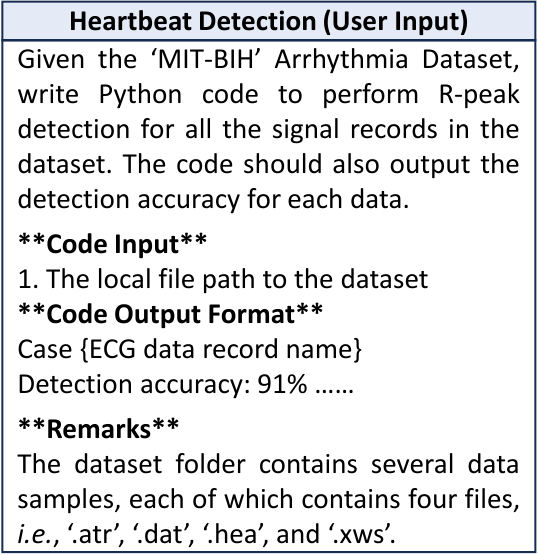}
        \label{fig:problem_statement_HD}
    }
    \centering
    \subfigure[Problem statement for HAR]{
        \includegraphics[width=0.225\textwidth]{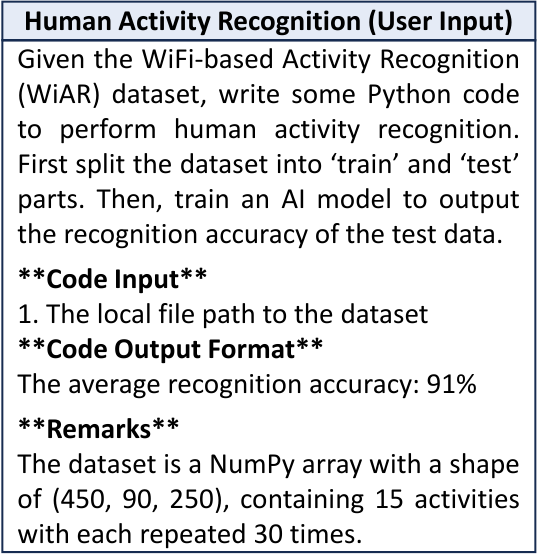}
        \label{fig:problem_statement_HAR}
    }
    \caption{Problem statements of the two applications (HAR and multimodal HAR share a similar prompt).}
    \label{fig:problem_statement}
\vspace{-15pt}
\end{figure}

We access GPT-4o and DeepSeek-Coder via API keys, interact with GitHub Copilot via Visual Studio Code's chat window, and deploy the rest on an edge server. Note that to our best, currently there is no LLM-based program synthesis system tailored for IoT application development. Therefore, we choose some SOTA code generation LLMs and systems as our baselines.

\vspace{-8pt}
\subsection{Application Evaluation}
\vspace{-3pt}
With the designed two problem statements (Fig.~\ref{fig:problem_statement}) for the three IoT applications, we input them into \project and the baselines to synthesize 20 different programs for each task.
We then evaluate their performance based on the metrics described in \S~\ref{sec:overall_metric}.

\vspace{-3pt}
\subsubsection{\textbf{HD}}
\begin{figure}
\vspace{-8pt}
\setlength{\abovecaptionskip}{-5pt}
        \subfigtopskip=-6pt
        \subfigcapskip=-6pt
    \centering
        \subfigure[Precision]{
            \includegraphics[width=0.225\textwidth]{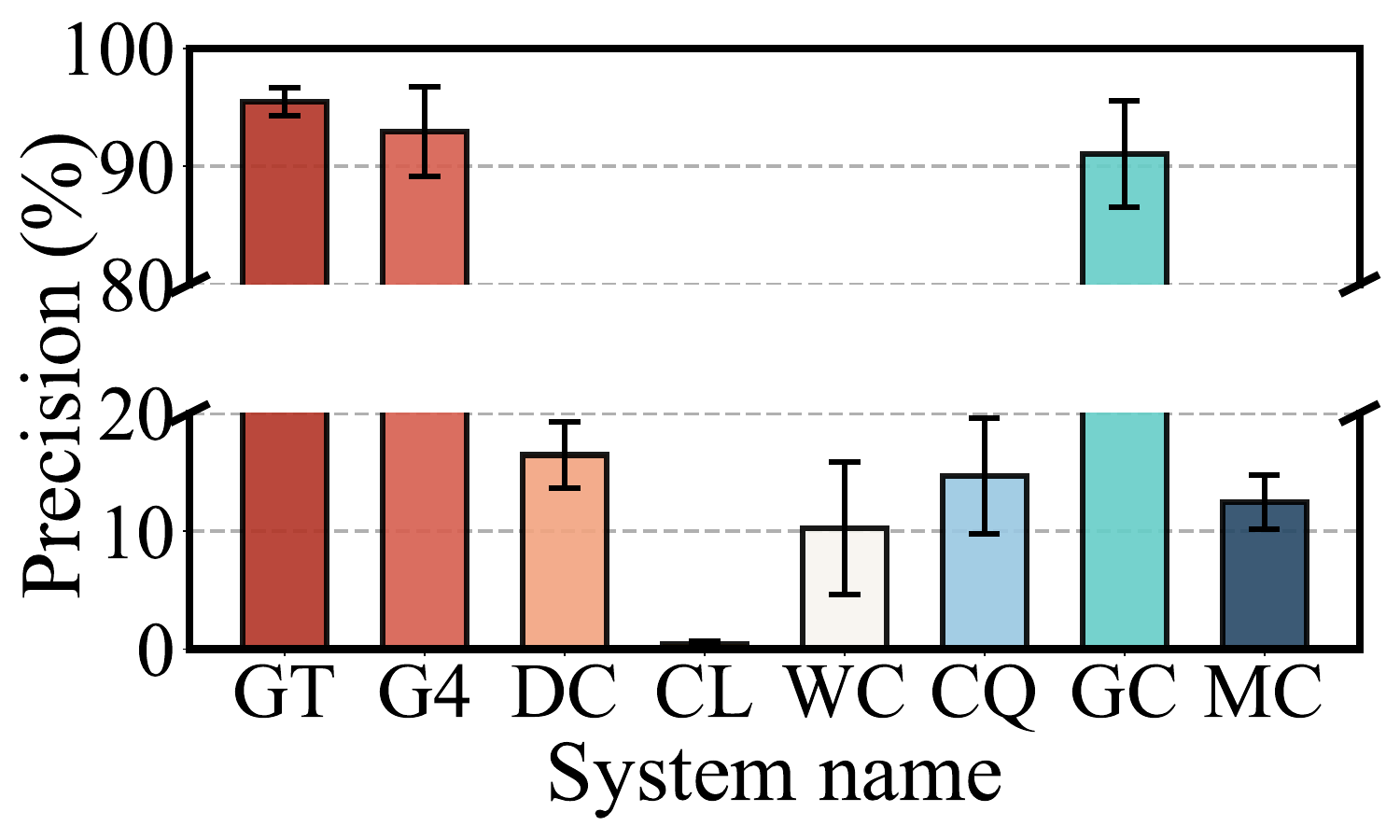}
            \label{fig:precision_rate_HD}
        }
        \centering
        \subfigure[Recall rate]{
            \includegraphics[width=0.225\textwidth]{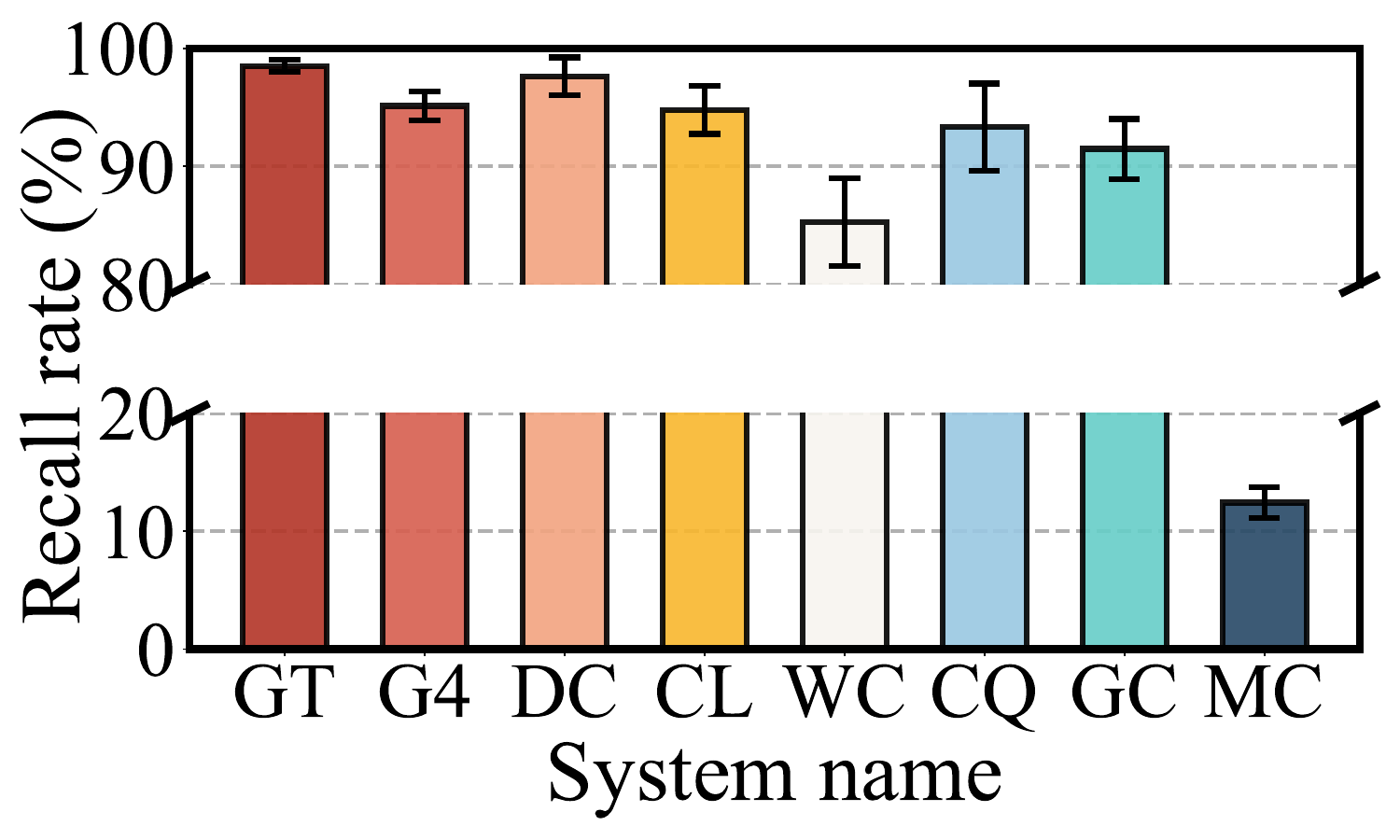}
            \label{fig:recall_rate_HAR}
        }
        \caption{The overall performance of HD.}
        \label{fig:overall_performance_HD}
    \vspace{-12pt}
\end{figure}

As shown in Fig.~\ref{fig:overall_performance_HD}, the code generated by \textit{GPIoT} significantly outperforms all the baselines, with an average precision gain of 64.7\% and an average RR increase of 16.9\%. It's also worth noting that CodeLlama, WizardCoder, and CodeQwen achieve moderate RR (above 80\%) but exhibit lower precision. With further analysis of the code, we find that they all adopt a simple peak detection function, \verb|scipy.signal.find_peaks()|, which typically fails when handling abnormal ECG data from patients. As a result, the detection results contain numerous false positives with low precision. Additionally, after reviewing the code generated by MapCoder, we observe that it incorporates more advanced heartbeat signal processing algorithms (\eg, bandpass filtering and adaptative thresholding). However, the final program exhibits a significant performance drop compared with other baselines. This is because heartbeat detection is a relatively simple IoT application that does not require highly sophisticated planning and iterative debugging. Integrating many advanced algorithms into a simple signal-processing program may lead to inconsistency issues. In other words, the heartbeat signal may be over-processed by these algorithms, leading to degraded performance. \cite{tejedor2019multiple, han2022real}. On the contrary, the code generated by \textit{GPIoT} utilizes dedicated algorithms (\eg, Pan-Tompkins) for R-peak detection due to embedded IoT domain knowledge during tuning, consistently achieving high precision and RR.

\vspace{-3pt}
\subsubsection{\textbf{HAR}}
\begin{figure}
\vspace{6pt}
\setlength{\abovecaptionskip}{-5pt}
        \subfigtopskip=-6pt
        \subfigcapskip=-6pt
    \centering
        \subfigure[Classification accuracy]{
            \includegraphics[width=0.225\textwidth]{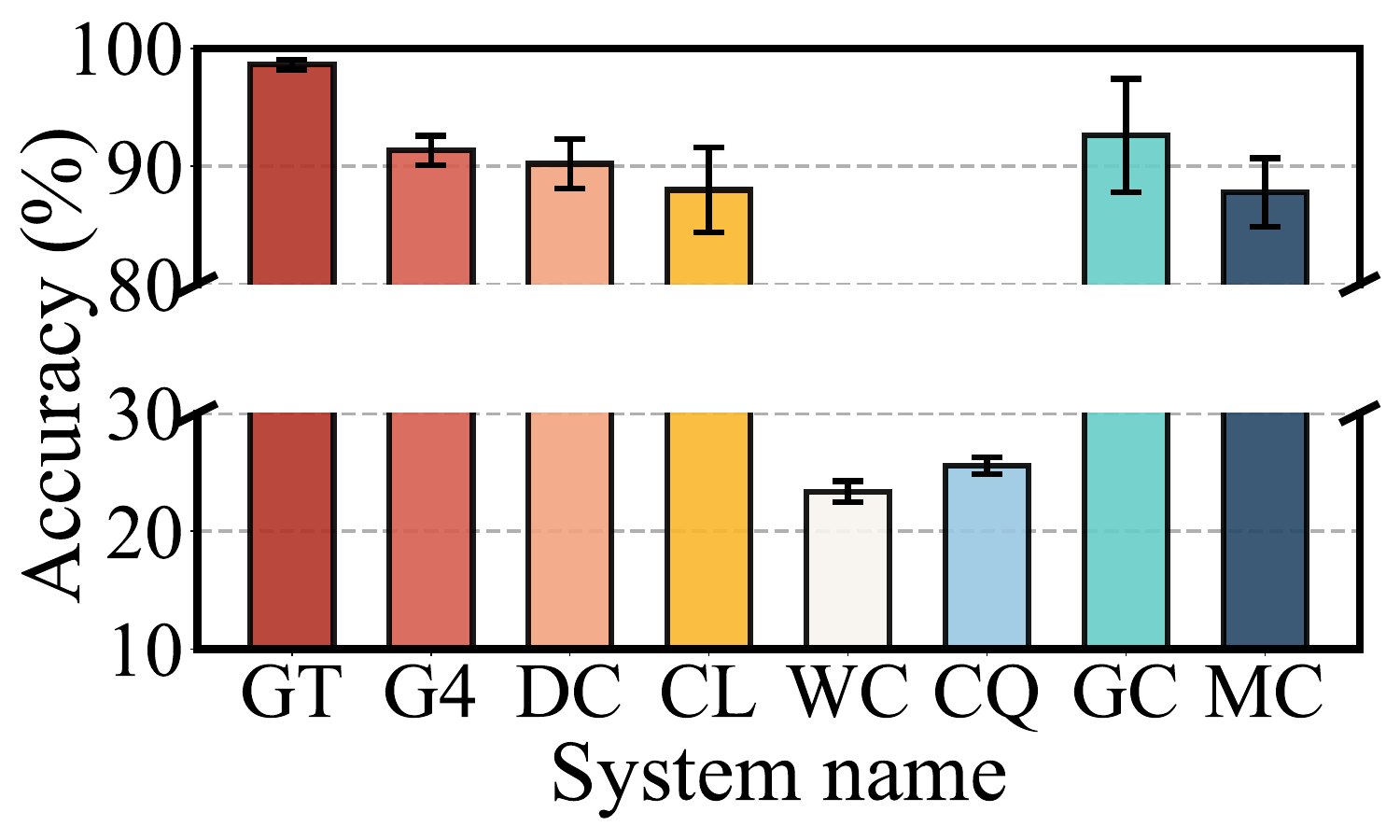}
            \label{fig:classification_accuracy_HAR}
        }
        \centering
        \subfigure[GPU memory \& inference time]{
            \includegraphics[width=0.225\textwidth]{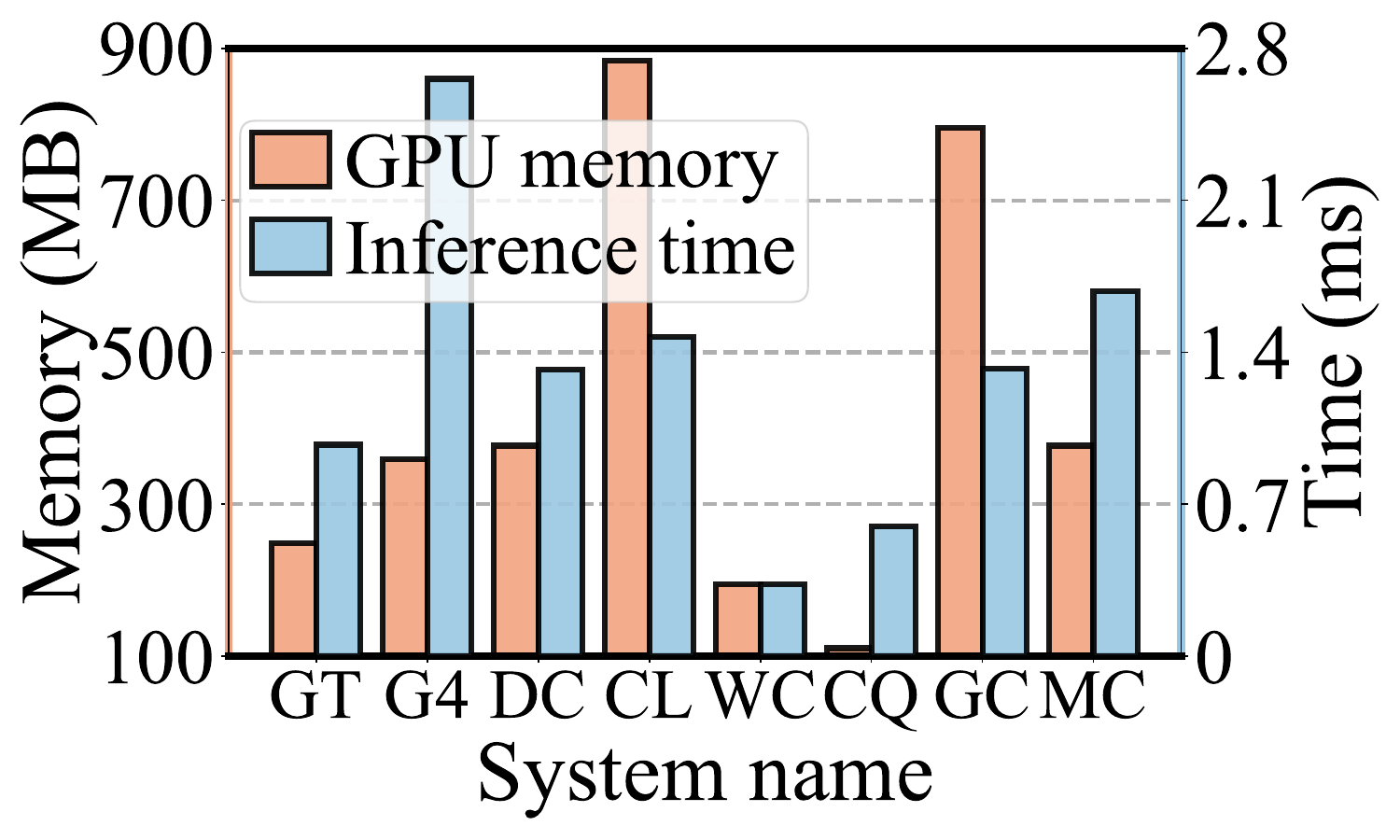}
            \label{fig:memory_time_HAR}
        }
        \caption{The overall performance of HAR.}
        \label{fig:overall_performance_HAR}
    \vspace{-18pt}
\end{figure}

In this evaluation, for all the generated HAR models, we set the training epochs to 10 and the batch size to 32 for a fair comparison. Besides, during our implementation, we find that after around 15 training epochs, all the models gradually converge. Therefore, we compare the model performance at the 10th epoch. As shown in Fig.~\ref{fig:classification_accuracy_HAR}, the program synthesized by \textit{GPIoT} achieves a 17.2\% higher accuracy with 47.8\% less GPU memory and 38.3\% shorter inference time on average. By analyzing the generated code, we find \textit{GPIoT} applies: 1) a data preprocessing method, Butterworth low-pass filtering, on the WiFi data, considering that the low-frequency components of WiFi CSI are primarily influenced by human activities \cite{wang2015understanding}. 2) an augmentation method tailored for IoT data (\eg, time-frequency masking \cite{zhouLearningRingsSelfSupervised2022}) on the WiFi signal to further enhance the diversity of the dataset. 3) a CUDA optimization mechanism \cite{choi2021implementing} to reduce GPU memory usage while enhancing runtime efficiency. In contrast, the baselines directly input raw WiFi data into HAR models without GPU optimization, leading to diminished performance and heightened memory consumption. Moreover, the error bar of \project is smaller than that of the baselines, indicating that \project generates more stable responses with a more robust performance of the synthesized program. Note that MC can synthesize programs with competitive performance since HAR is a more complex application. Such advancements of \textit{GPIoT} originate from our meticulously crafted IoT-specialized datasets and the PECT paradigm, which embeds abundant IoT domain knowledge from our datasets into the tuned SLMs for more consistent outputs.

Additionally, considering that our HAR model is deployed on edge devices with various resource constraints \cite{lin2023workload, shen2024fedconv, shen2023feddm, wang2024swapnet, wang2024latte}, we involve different resource requirements in the prompts for the generated code to facilitate a more comprehensive evaluation. Specifically, we first design a base prompt: "I need to deploy the HAR model on Jetson Nano" (\textbf{P1}) without any resource specifications. Based on \textbf{P1}, we then create three variations by adding different resource constraints: "Do not consider any resource constraints but only model accuracy" (\textbf{P2}), "The GPU memory usage should not exceed 200 MB" (\textbf{P3}), and "The GPU memory should not exceed 50 MB" (\textbf{P4}). We then instruct \project to synthesize 20 different versions of programs using each of the prompts above. After executing the programs with the same configurations, we record the average classification accuracy and GPU memory usage, as shown in Fig~\ref{fig:resource_constrained_prompt}. We find that: 1) Given P2, \project can construct an AI model with a large number of parameters to achieve ultimate performance, with its classification accuracy approaching nearly 100\%. 2) Given P3, \project can adopt a smaller model within the resource budget (\ie, GPU memory usage not exceeding 200 MB) with a slight performance drop. 3) Given P4, \project employs a highly optimized model requiring only approximately 50 MB of GPU memory, resulting in an acceptable accuracy drop of 5\%. These results demonstrate that \project can generate code tailored to different resource budgets, stemming from our \textit{IoT-Oriented Data Augmentation} method, which augments data samples considering resource heterogeneity of target devices in IoT applications.

\begin{figure}
    \vspace{-8pt}
\setlength{\abovecaptionskip}{-5pt}
        \subfigtopskip=-6pt
        \subfigcapskip=-6pt
   \centering
    \subfigure[Classification accuracy]{
        \includegraphics[width=0.225\textwidth]{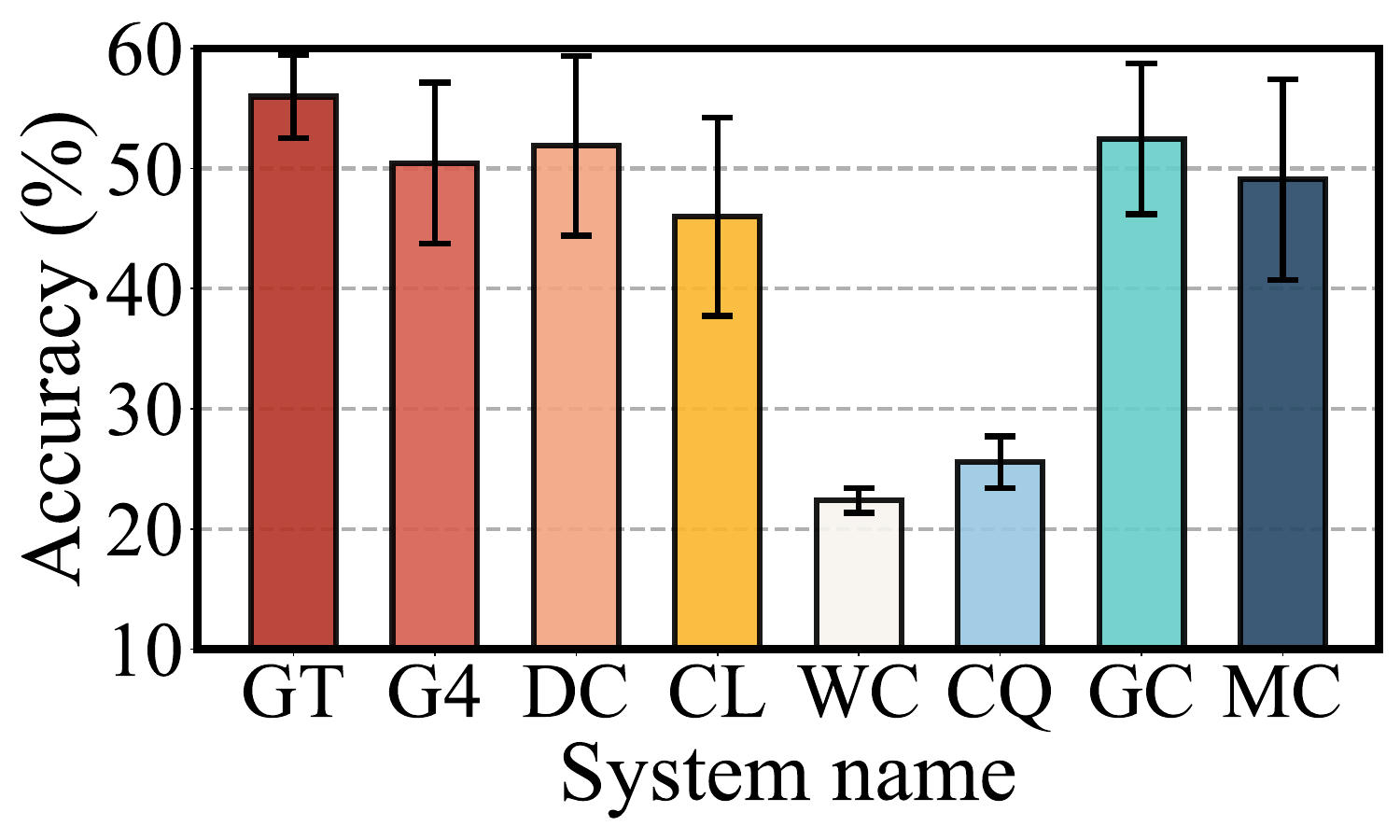}
        \label{fig:overall_multimodal_accuracy}
    }
    \centering
    \subfigure[GPU memory \& inference time]{
        \includegraphics[width=0.225\textwidth]{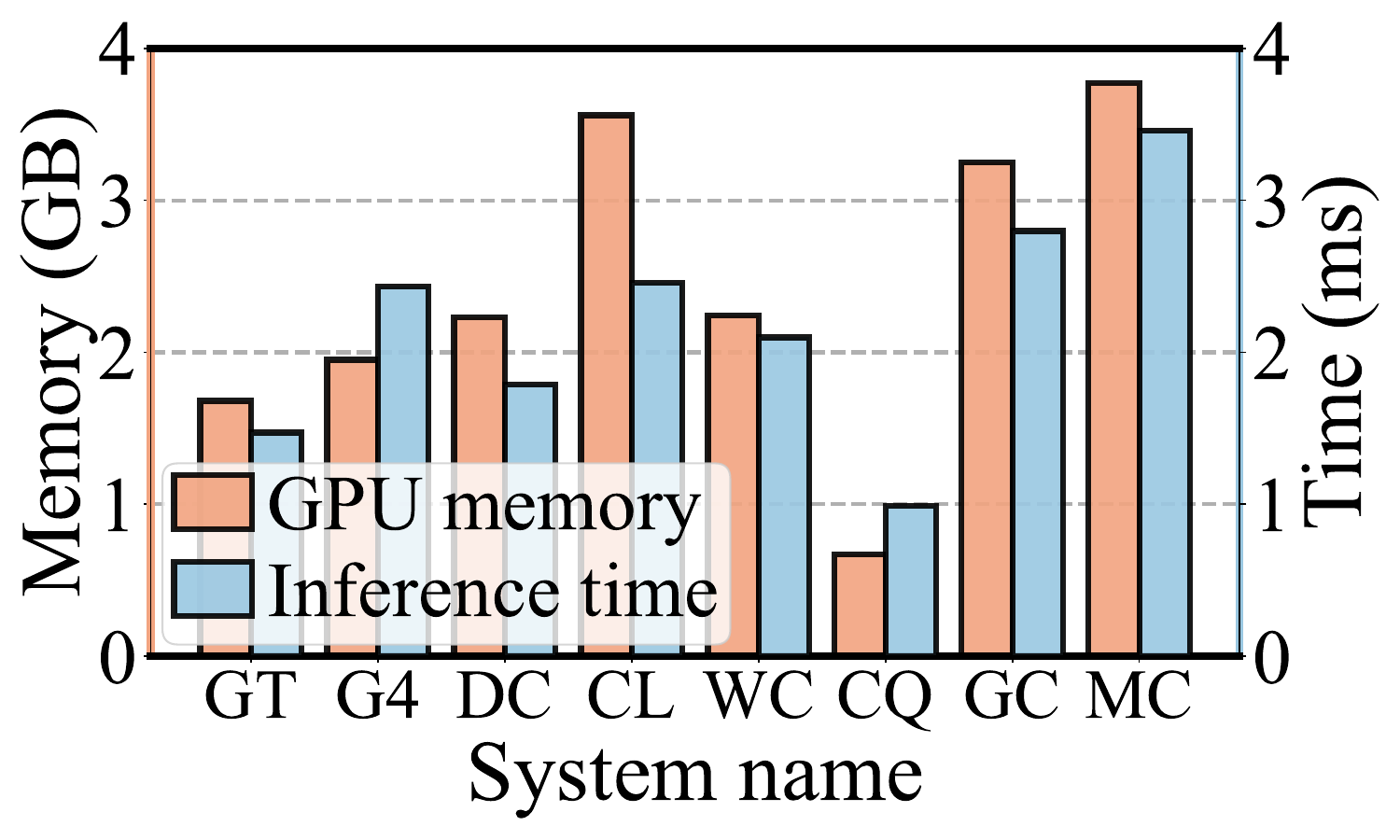}
        \label{fig:overall_multimodal_memory}
    }
    \caption{The overall performance of multimodal HAR}
    \label{fig:overall_multimodal}
    \vspace{-14pt}
\end{figure}

\begin{figure}[t]
\vspace{4pt}
\setlength{\abovecaptionskip}{-5pt}
        \subfigtopskip=-6pt
        \subfigcapskip=-6pt
    \centering
        \subfigure[The BLEU score]{
            \includegraphics[width=0.225\textwidth]{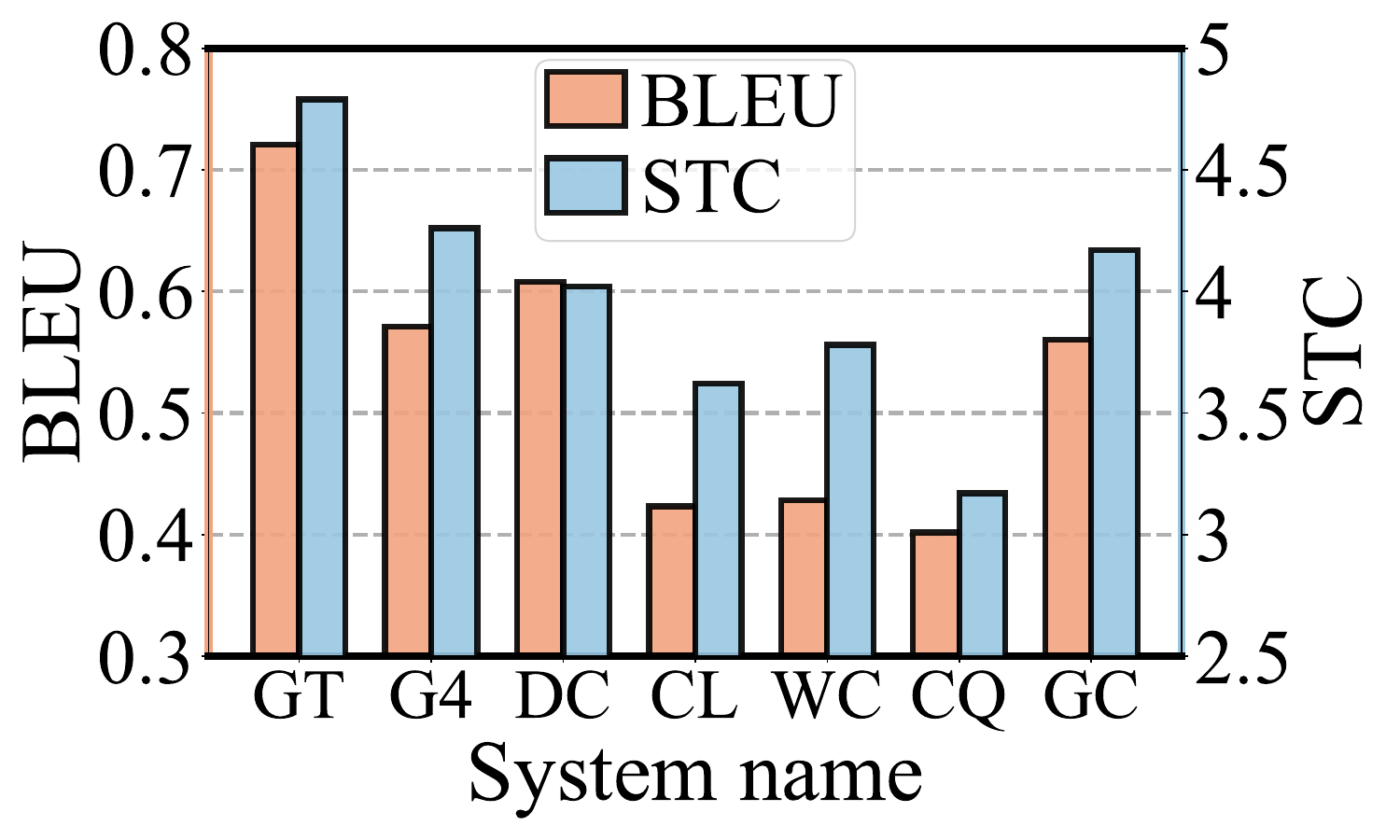}
            \label{fig:task_decomposition_bleu}
        }
        \centering
        \subfigure[The Format Correctness Rate]{
            \includegraphics[width=0.225\textwidth]{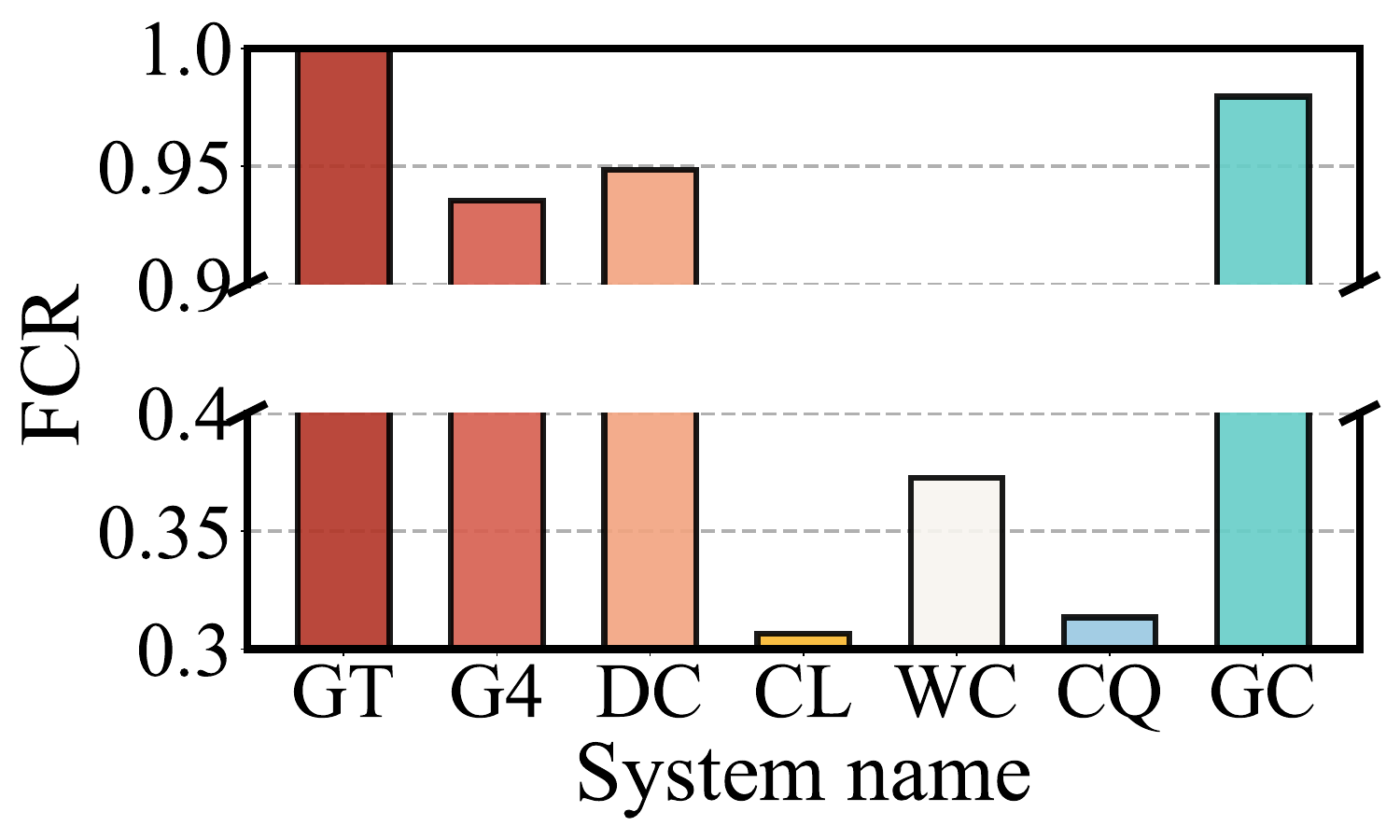}
            \label{fig:task_decomposition_FCR}
        }
        \caption{Breakdown evaluation on TDSLM.}
        \label{fig:task_decomposition_SLM_evaluation}
    \vspace{-18pt}
\end{figure}

\subsubsection{\textbf{Multimodal HAR}}
We further instruct \project to synthesize programs for the multimodal HAR application, aiming to evaluate its programming ability for more complex tasks. As shown in Fig.~\ref{fig:overall_multimodal}, compared with the baselines, the program synthesized by \project achieves an average accuracy improvement of 13.44\% while requiring moderate GPU memory and inference time. After reviewing the source code, we find that both \project and the baselines train three encoders to first extract useful features from different modalities, followed by a classifier to recognize the corresponding activity. However, the program synthesized by \project adopts some model optimization methods (\eg, quantization or pruning) and data augmentation methods tailored for IoT sensor data (\eg, time-frequency masking). As such, the synthesized program can train a memory-optimized model while maintaining high classification accuracy. These results indicate that, benefiting from our SLM tuning, the program synthesized by \project can incorporate more IoT-specific data processing and model optimization algorithms, thereby achieving high performance even for multimodal HAR.

\vspace{-5pt}
\subsection{Breakdown Evaluation}
\vspace{-3pt}
We separately evaluate TDSLM and CGSLM on \textit{\textbf{IoTBench}} to explore the effectiveness of fine-tuning in the IoT domain.


\vspace{-3pt}
\subsubsection{\textbf{Metrics}} We adopt different metrics for the two SLMs.

\noindent\textbf{TDSLM.} 1) \textit{BLEU}: we measure the BLEU score \cite{papineni2002bleu} between the generated decomposed tasks and the
reference from \textit{IoTBench}.
A larger BLEU score indicates higher semantic similarity and higher task decomposition quality. 2) \textit{Format Correctness Rate} (FCR): the portion of TDSLM's outputs that correctly separate each decomposed task with a blank line for the convenience of further processing.
This aims to quantify TDSLM's instruction-following and text-formatting abilities. 3) \textit{Sub-Task Completeness} (STC): we invite 10 IoT experts to assess the extent to which the decomposed tasks cover all essential parts of the application based on the reference.

\noindent\textbf{CGSLM.} 1) \textit{Code embedding similarity}: we use CodeT5+ \cite{wang2023codet5+} to convert code snippets into embeddings and compute the cosine similarity between embeddings of the generated and reference code. A higher similarity indicates a stronger ability to generate IoT-related code. 2) \textit{Pass@k}: we measure the pass@k value by calculating the portion of programs that pass all the test cases. A higher value indicates better performance of the generated code. 3) \textit{User Requirement Coverage} (URC): we first ask the users to execute the generated code and review the documentation. Next, they are asked to evaluate the extent to which the generated code and documentation fulfill all the user requirements. 4) \textit{Code quality}: we assess the quality of the generated code by adopting a commercial-off-the-shelf (COTS) code quality verification tool, SonarQube \cite{sonar}, detecting bug/logic errors, security issues, and code smells \cite{sonarsource}. Code smells are not bugs but bad coding styles (\eg, variable name mismatching regular expression) or potential weaknesses (\eg, package version incompatibility).
Note that STC and URC are user-related metrics, which are rated on a scale from 1 (not at all) to 5 (completely).

\begin{figure}[t]
\vspace{-8pt}
\setlength{\abovecaptionskip}{-5pt}
        \subfigtopskip=-6pt
        \subfigcapskip=-6pt
    \centering
        \subfigure[The embedding similarity]{
            \includegraphics[width=0.225\textwidth]{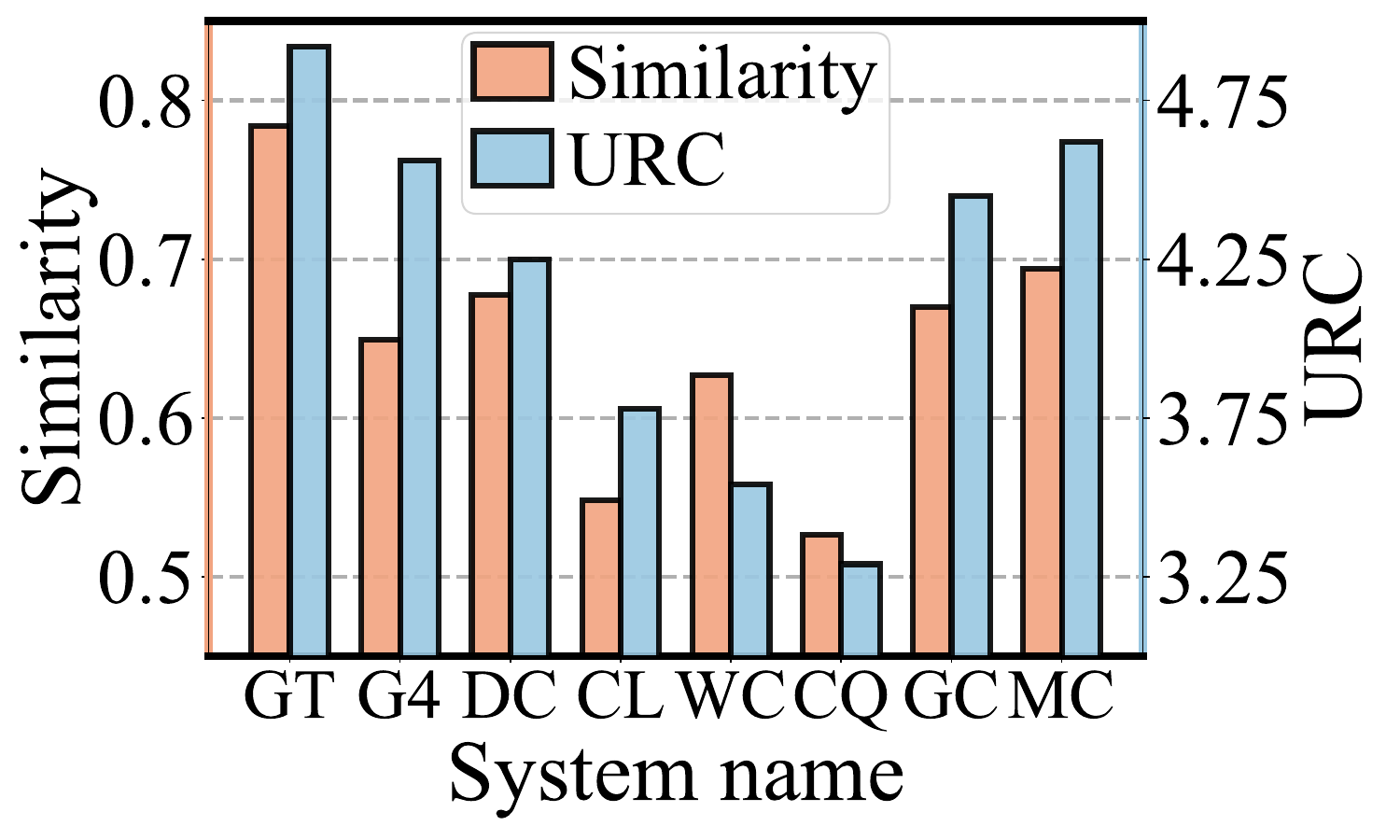}
            \label{fig:code_generation_bleu}
        }
        \centering
        \subfigure[The pass@k]{
            \includegraphics[width=0.225\textwidth]{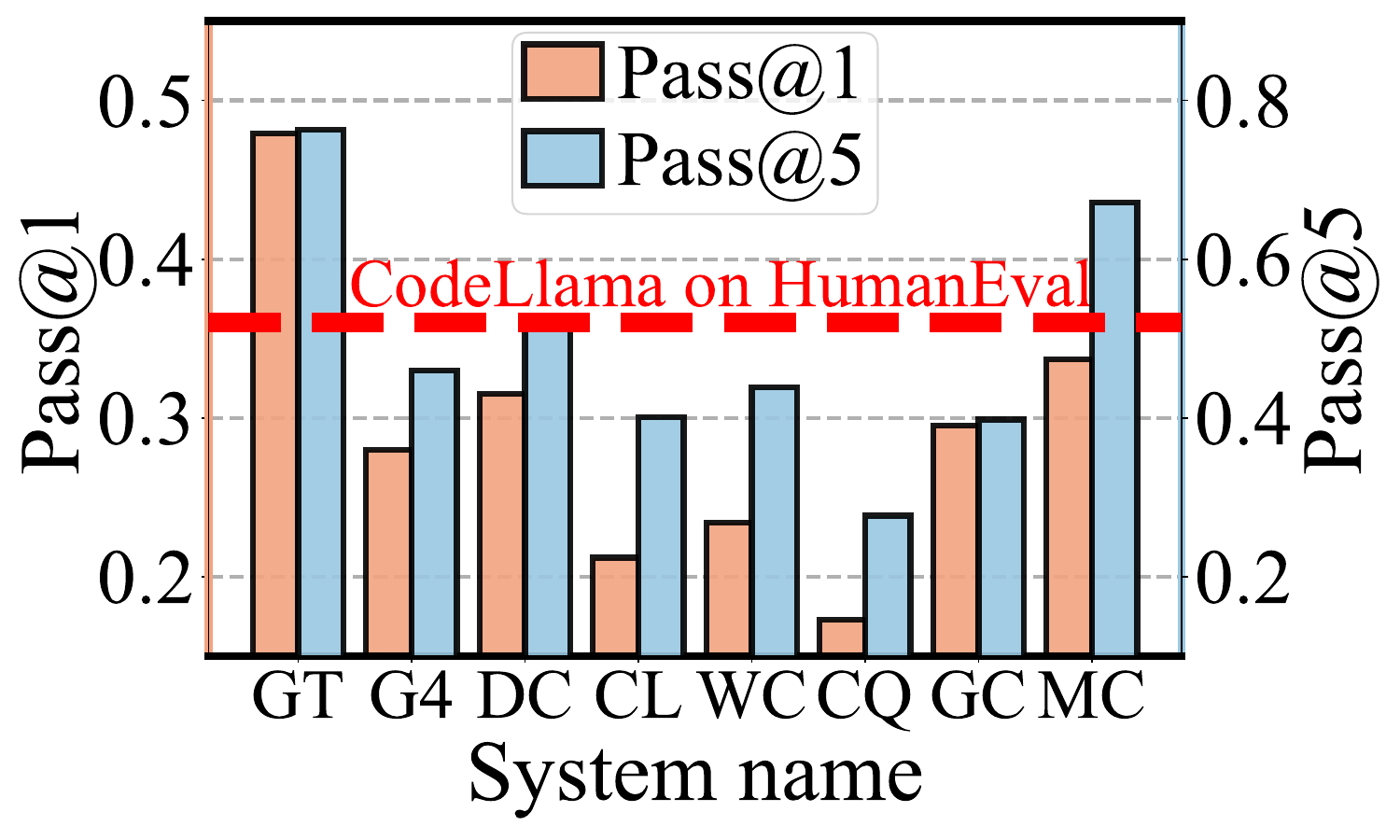}
            \label{fig:code_generation_pass_k}
        }
        \caption{Breakdown evaluation on CGSLM}
        \label{fig:code_generation_SLM_evaluation}
    \vspace{-14pt}
\end{figure}

\begin{figure}
\vspace{4pt}
\setlength{\abovecaptionskip}{-5pt}
        \subfigtopskip=-6pt
        \subfigcapskip=-6pt
   \centering
    \subfigure[Specifying resource constraints]{
        \includegraphics[width=0.225\textwidth]{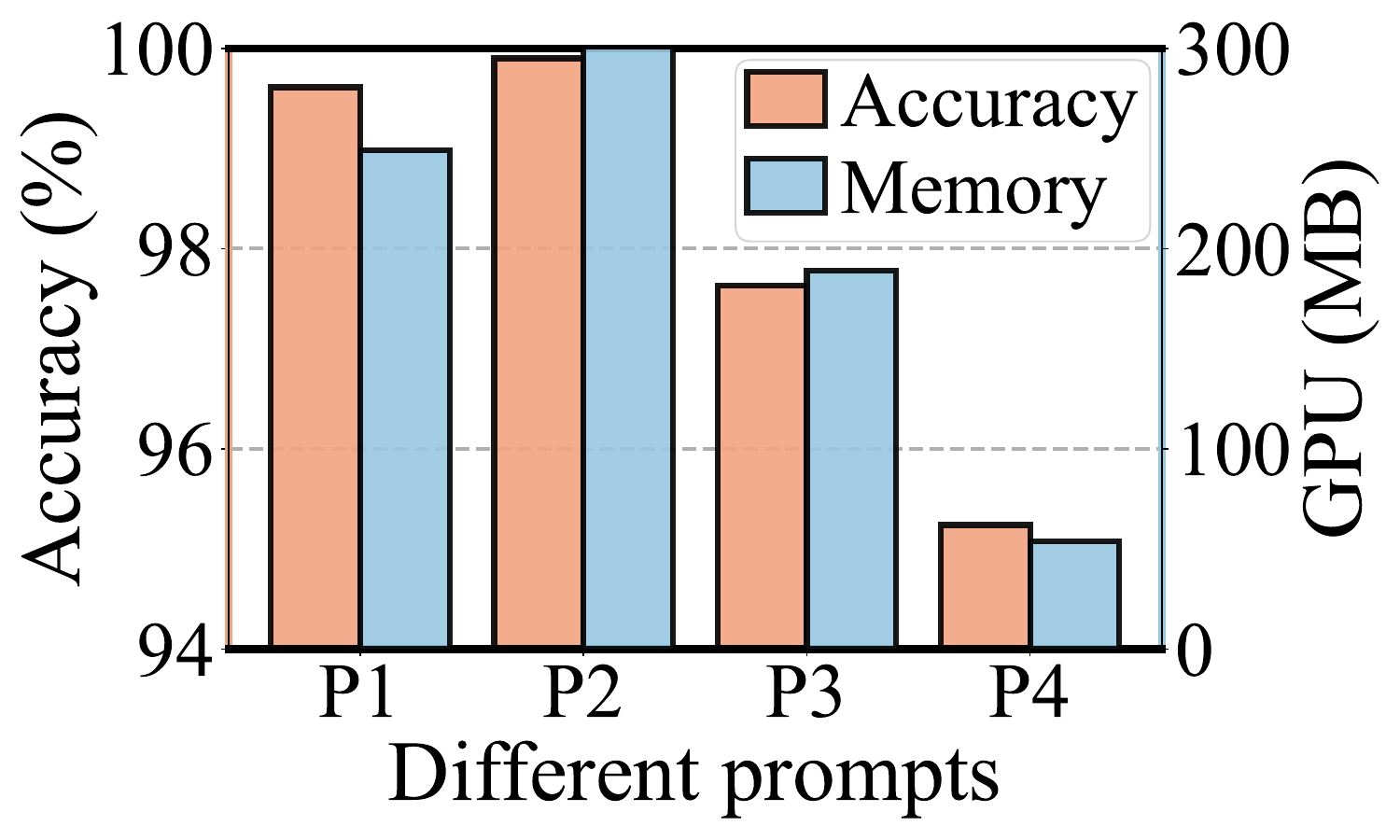}
        \label{fig:resource_constrained_prompt}
    }
    \centering
    \subfigure[Code quality evaluation]{
        \includegraphics[width=0.225\textwidth]{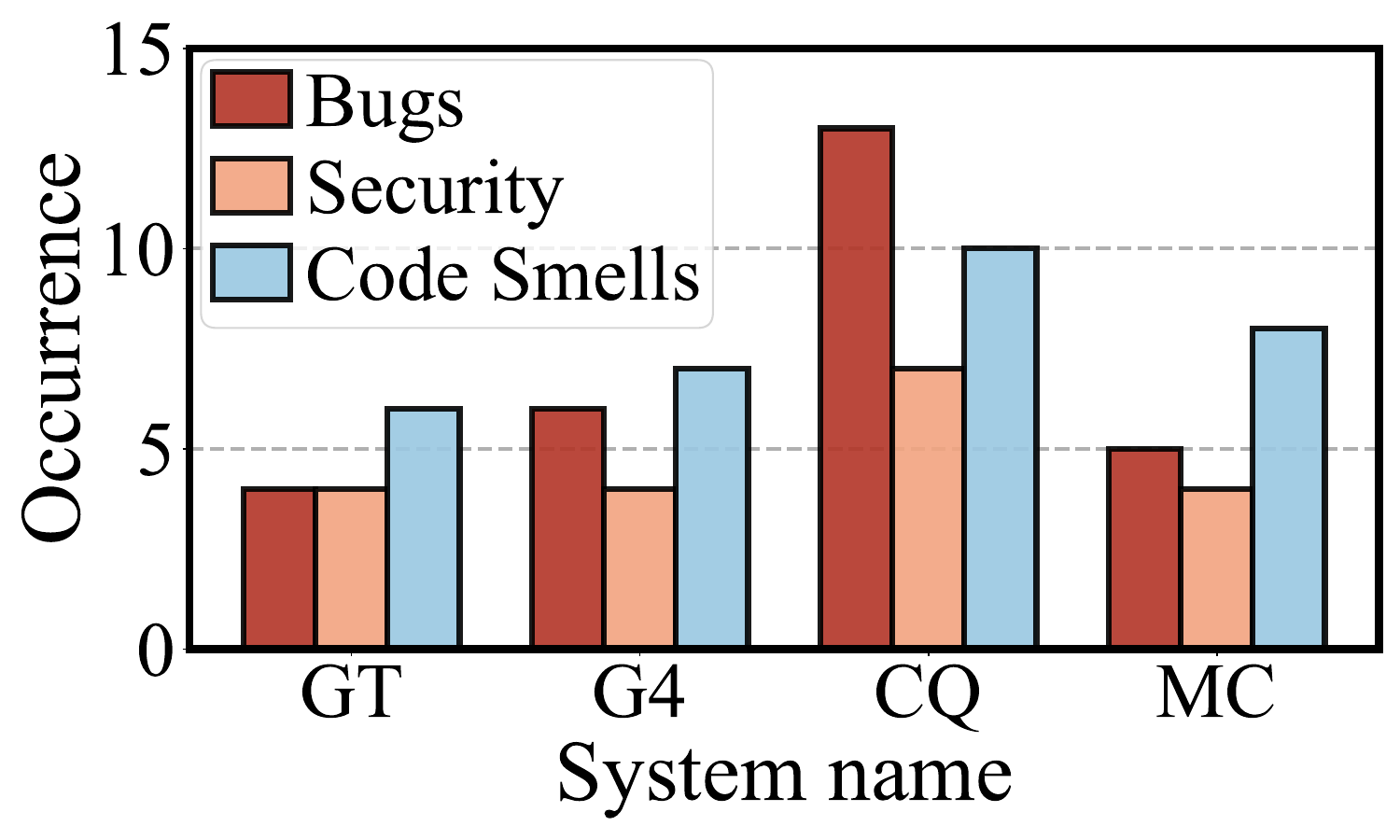}
        \label{fig:code_quality}
    }
    \caption{(a) Evaluating \project's performance using prompts with different resource constraints. (b) Evaluating the quality of the code generated by CGSLM using SonarQube.}
    \vspace{-18pt}
\end{figure}

\vspace{-3pt}
\subsubsection{\textbf{TDSLM}}
\label{sec:break_down_task_decomposition}
We input each \textit{problem statement} from \textit{IoTBench} into TDSLM and the baselines to generate 20 different \textit{decomposed tasks} and calculate the average BLEU score, FCR, and URC. From Fig.~\ref{fig:task_decomposition_SLM_evaluation}, we observe: 1) The decomposed tasks generated by TDSLM achieve a 48\% higher BLEU score than the baselines on average, indicating a stronger decomposition ability for IoT tasks. 2) TDSLM achieves 99\% FCR, indicating remarkable stability to generate intermediate output (\textit{decomposed tasks}) based on pre-defined formats. 3) TDSLM also achieves a 28\% higher STC on average, showcasing strong abilities in understanding IoT knowledge and generating comprehensive decomposed tasks for IoT applications. Such superior IoT task decomposition and data formatting performance of TDSLM originate from the tuning process on TDD with our IoT-oriented text data augmentation method.


\vspace{-3pt}
\subsubsection{\textbf{CGSLM}}
\label{sec:pass_k_metri}
We input each \textit{task specification} from \textit{IoTBench} into CGSLM and the baselines to generate 20 different \textit{code \& documentation}. We then report the average code embedding similarity, pass@1, pass@5, URC, and the number of various issues detected by SonarQube. by executing and reviewing the code. For comparison, we also report the pass@1 achieved by CodeLlama on a general-purpose programming benchmark, HumanEval \cite{chen2021evaluating}. From Fig.~\ref{fig:code_generation_SLM_evaluation}, we observe that: 1) 1) CGSLM achieves an 18\% higher code embedding similarity than the baselines on average, implying stronger capability and generalizability in generating IoT-related code snippets. 2) CGSLM achieves higher pass@1 and pass@5 than the baselines, with an average increase of 21.5\% and 31\%, respectively, showcasing higher quality and accuracy of the generated code for solving IoT tasks. 3) Nearly all the baselines achieve much lower pass@1 for IoT programming tasks than CodeLlama on HumanEval due to limited capabilities in the IoT domain. 4) Users show a stronger preference for CGSLM over the baselines, with an average increase of 23\% in URC. 5) The program synthesized by CGSLM contains fewer bugs, security issues, and code-smell-related issues. This is primarily because, CGSLM, fine-tuned on our manually crafted datasets, can synthesize programs with enhanced code quality. Such superior abilities of CGSLM in generating IoT-related code snippets with high URC stems from our co-tuning on CGD with well-structured data. Consequently, CGSLM can generate code adopting more IoT-specific algorithms following task specifications.


\begin{figure}[t]
\vspace{-8pt}
\setlength{\abovecaptionskip}{-5pt}
        \subfigtopskip=-6pt
        \subfigcapskip=-6pt
   \centering
    \subfigure[Impact on TDSLM]{
        \includegraphics[width=0.225\textwidth]{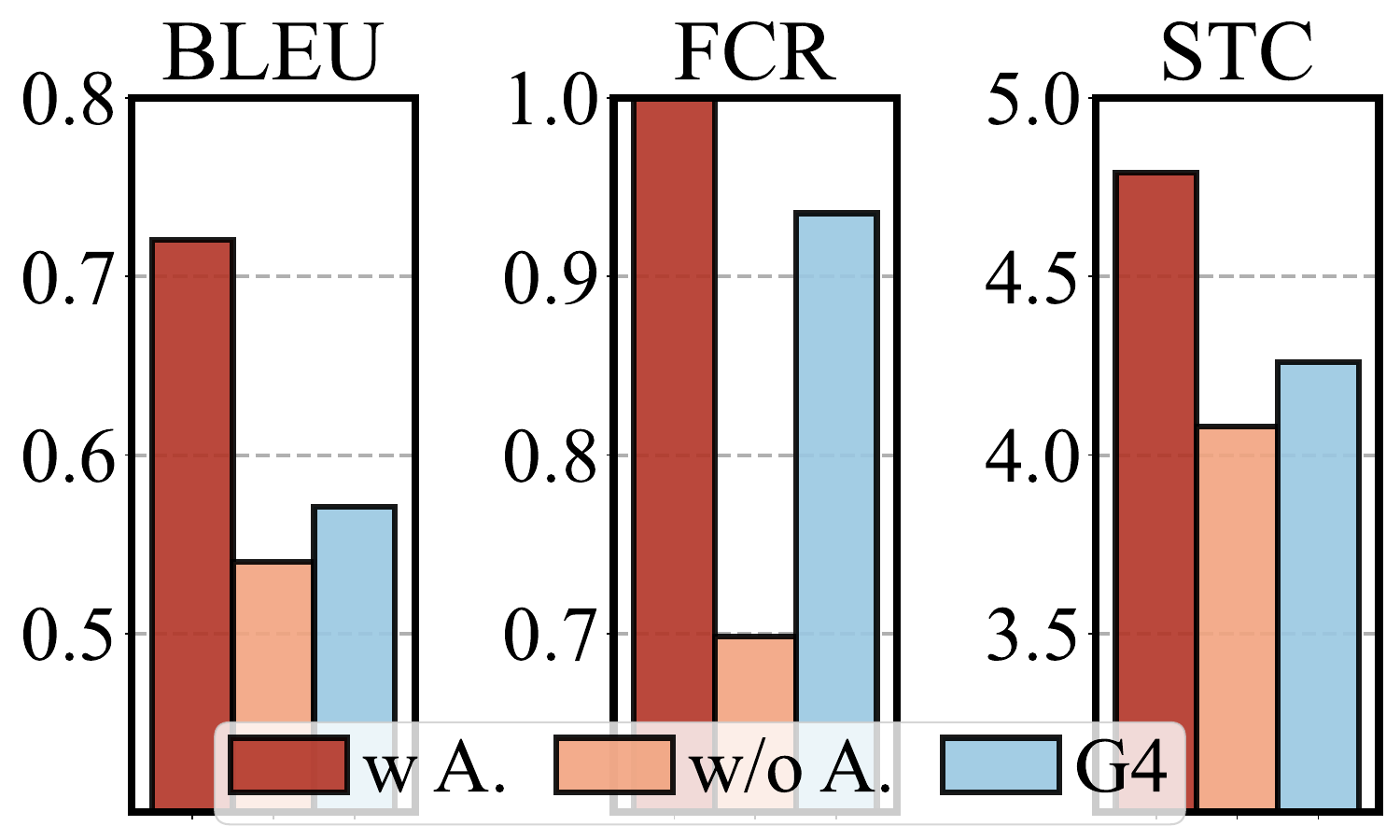}
        \label{fig:ablation_data_augmentation_single_module}
    }
    \centering
    \subfigure[Impact on HD]{
        \includegraphics[width=0.225\textwidth]{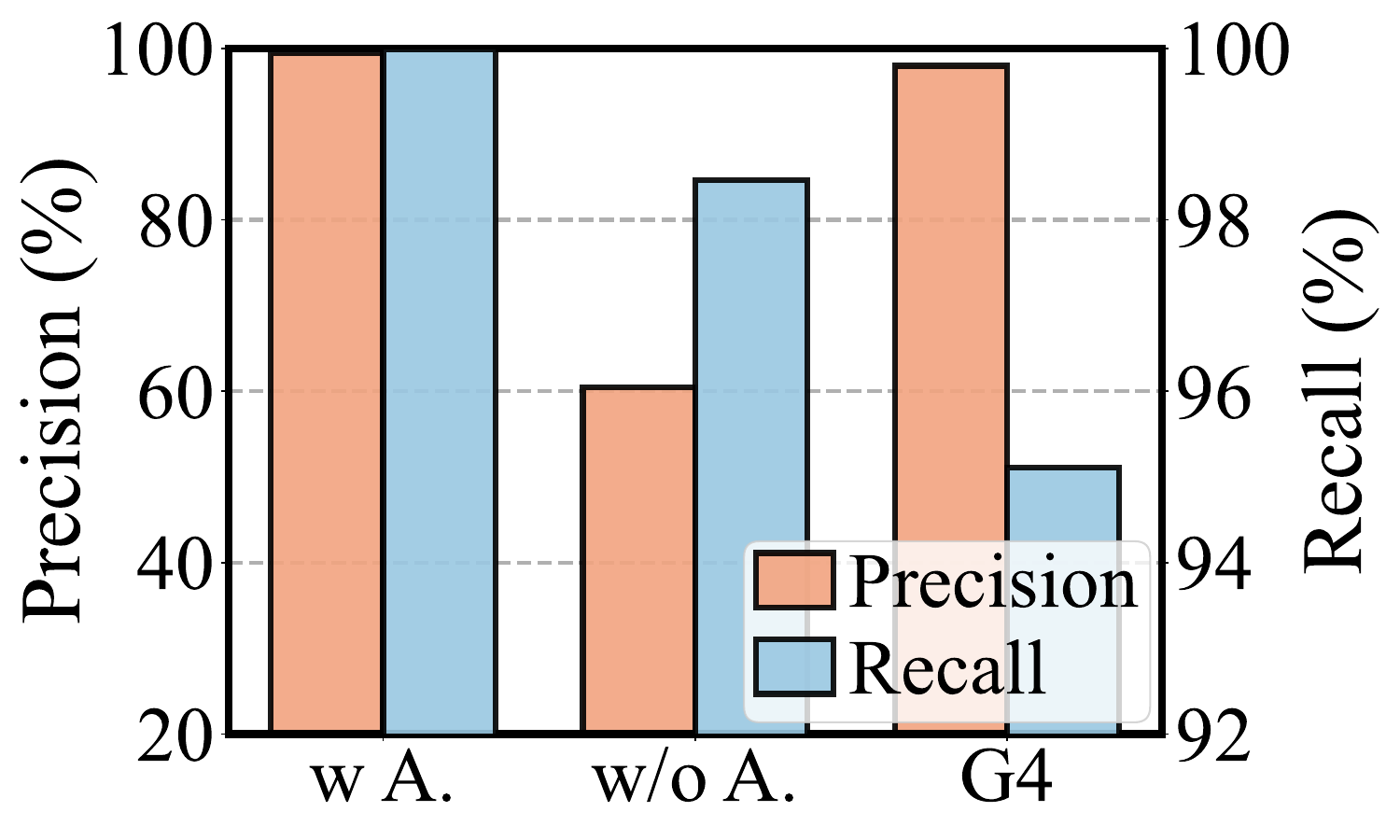}
        \label{fig:ablation_data_augmentation_whole_system}
    }
    \caption{Ablation of IoT-oriented augmentation (A.).}
    \label{fig:ablation_data_augmentation}
    \vspace{-12pt}
\end{figure}

\begin{figure}[t]
\vspace{4pt}
\setlength{\abovecaptionskip}{-5pt}
        \subfigtopskip=-6pt
        \subfigcapskip=-6pt
         \centering
    \subfigure[Impact on CGSLM]{
        \includegraphics[width=0.225\textwidth]{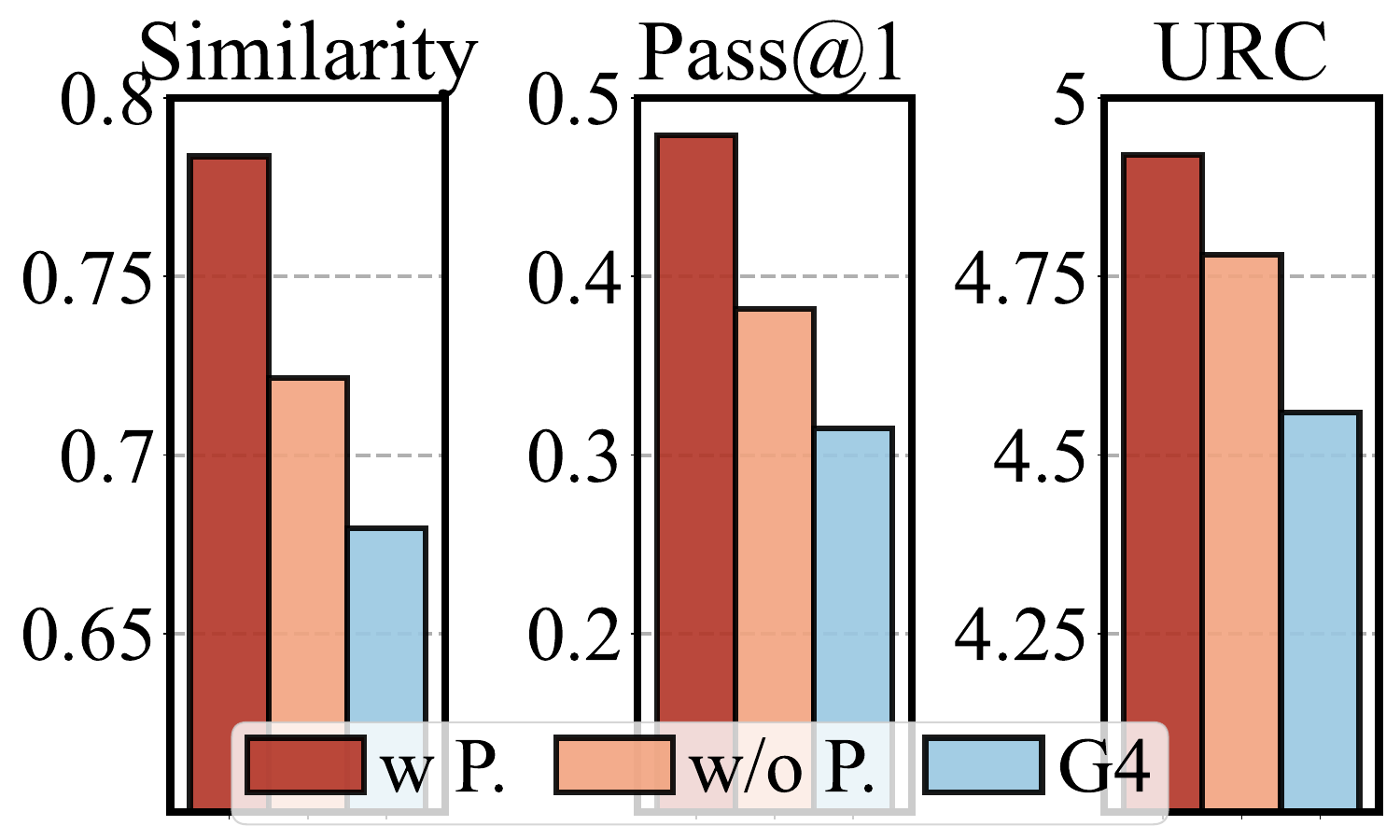}
        \label{fig:ablation_PECT_single_module}
    }
    \centering
    \subfigure[Impact on HAR]{
        \includegraphics[width=0.225\textwidth]{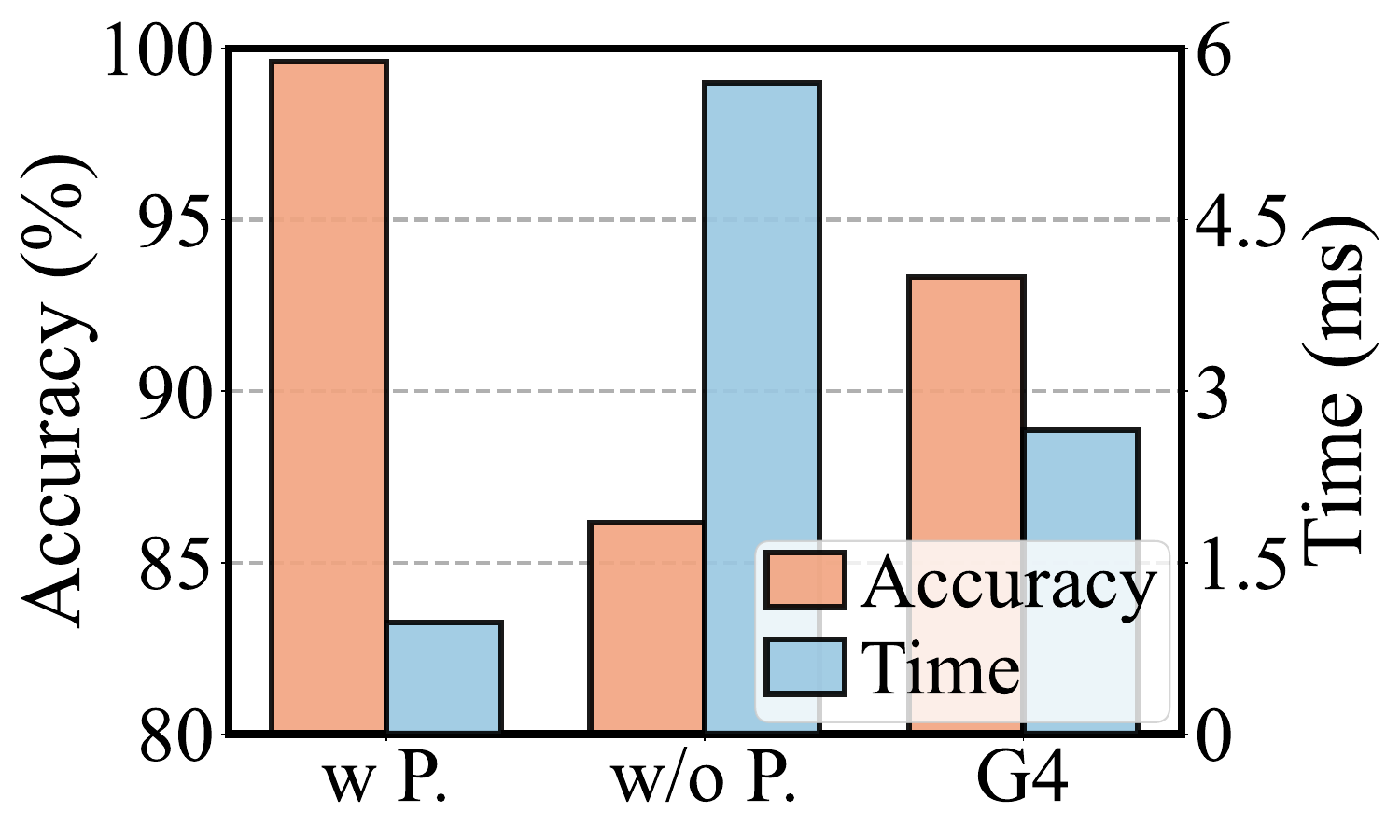}
        \label{fig:ablation_PECT_whole_system}
    }
    \caption{Ablation of PECT (P. in the figure).}
    \label{fig:ablation_PECT}
    \vspace{-18pt}
\end{figure}

\vspace{-8pt}
\subsection{Ablation Study}
\label{s:ablation}
\vspace{-2pt}
We conduct an ablation study by removing some proposed technical modules to investigate their importance to \textit{GPIoT}.

\vspace{-5pt}
\subsubsection{\textbf{IoT-Oriented Data Augmentation.}}
\label{sec:augmentation_ablation}
We directly tune TDSLM on the raw dataset without our IoT-oriented data augmentation, which contains only 273 data samples. We then evaluate the tuned model on \textit{IoTBench} and report the average BLUE score, FCR, and STC. As shown in Fig.~\ref{fig:ablation_data_augmentation_single_module}, without our augmentation, the tuned model exhibits a substantial performance decline across all metrics, much lower than GPT-4o. The main reason is that the raw dataset lacks generalizability and diversity in the IoT domain, which limits the tuned SLM's ability to decompose IoT problems into manageable sub-tasks, occasionally leading to incorrect results due to hallucinations. As a result, if we use such an insufficiently tuned TDSLM for heartbeat detection, \textit{GPIoT} still adopts a simple peak detection algorithm, leading to significant performance degradation of the generated code in both precision and recall rate (Fig.~\ref{fig:ablation_data_augmentation_whole_system}). The results demonstrate the importance of our IoT-oriented text data augmentation method in improving TDSLM's capability in task decomposition and IoT domain knowledge comprehension.


\vspace{-3pt}
\subsubsection{\textbf{PECT}}
\label{sec:PECT_ablation}
We separately fine-tune TDSLM and CGSLM on their own datasets without our PECT paradigm. We then evaluate the performance of the tuned CGSLM on \textit{IoTBench} and measure the average code embedding similarity, pass@1, and URC. As shown in Fig.~\ref{fig:ablation_PECT_single_module}, without PECT, the code generated by CGSLM exhibits performance degradation across all the metrics. This is because some IoT domain knowledge possessed by TDSLM cannot be shared with CGSLM. As a result, CGSLM cannot handle some programming tasks that are out of the scope, providing simple programs with degraded performance. However, even such insufficiently tuned CGSLM still outperforms GPT-4o, highlighting the advantage of fine-tuning in enhancing IoT-related code generation ability. Furthermore, without PECT in the HAR task, the final code neither adopts data pre-processing methods nor designs high-performance neural networks, leading to decreased classification accuracy, as shown in Fig.~\ref{fig:ablation_PECT_whole_system}. The results confirm the importance of our PECT paradigm in mitigating the domain misalignment issue and facilitating the IoT knowledge sharing between TDSLM and CGSLM.

\begin{figure}[t]
\vspace{-8pt}
\setlength{\abovecaptionskip}{-5pt}
        \subfigtopskip=-6pt
        \subfigcapskip=-6pt
   \centering
    \subfigure[Impact on HD]{
        \includegraphics[width=0.225\textwidth]{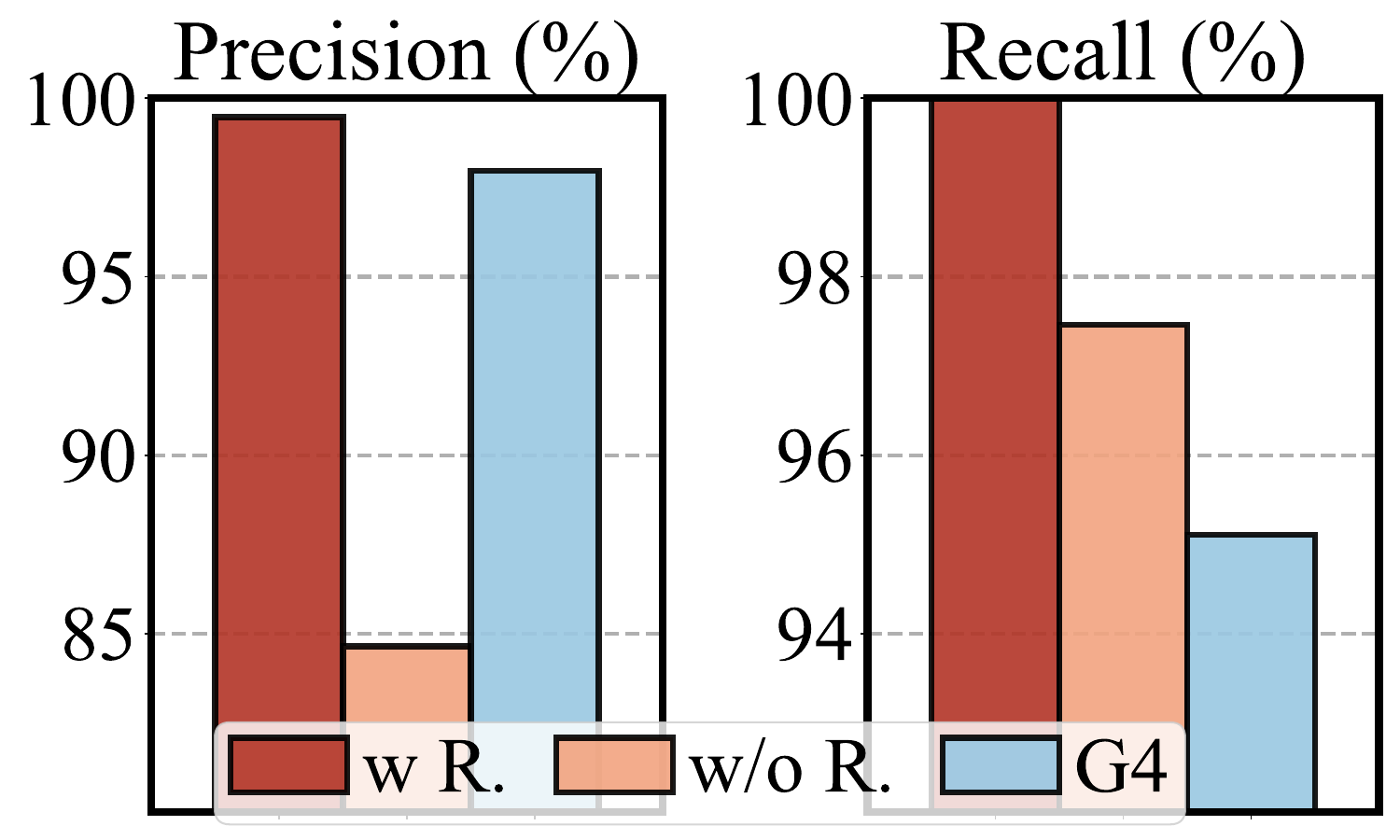}
        \label{fig:ablation_requirement_transformation_HD}
    }
    \centering
    \subfigure[Impact on HAR]{
        \includegraphics[width=0.225\textwidth]{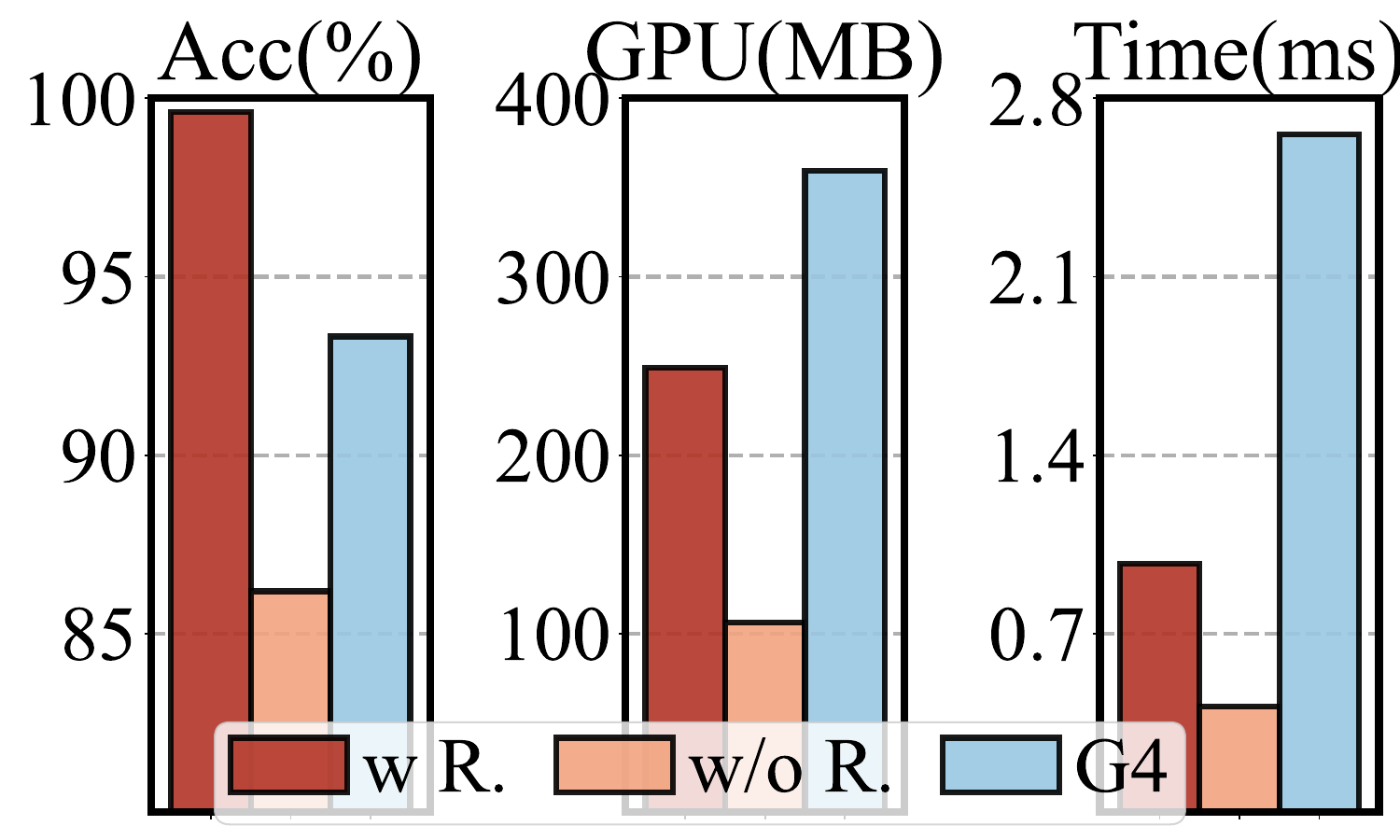}
        \label{fig:ablation_requirement_transformation_HAR}
    }
    \caption{Ablation of requirement transformation.}
    \label{fig:ablation_requirement_transformation}
    \vspace{-14pt}
\end{figure}

\begin{figure}
\vspace{8pt}
\setlength{\abovecaptionskip}{-5pt}
        \subfigtopskip=-6pt
        \subfigcapskip=-6pt
    \centering
    \subfigure[Signal processing-related tasks]{
        \includegraphics[width=0.2\textwidth]{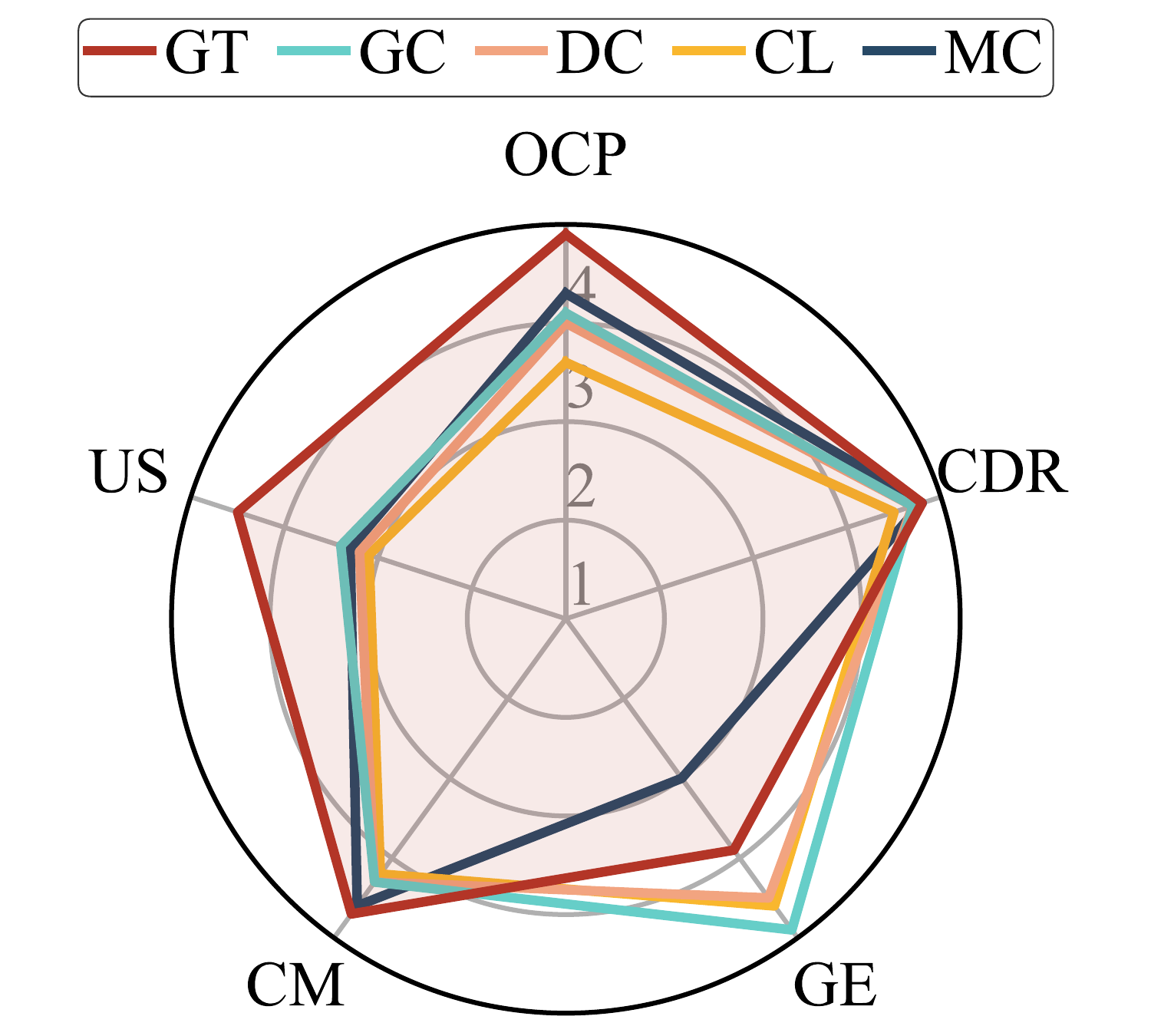}
        \label{fig:user_study_HD}
    }
    \centering
    \subfigure[Machine learning-related tasks]{
        \includegraphics[width=0.2\textwidth]{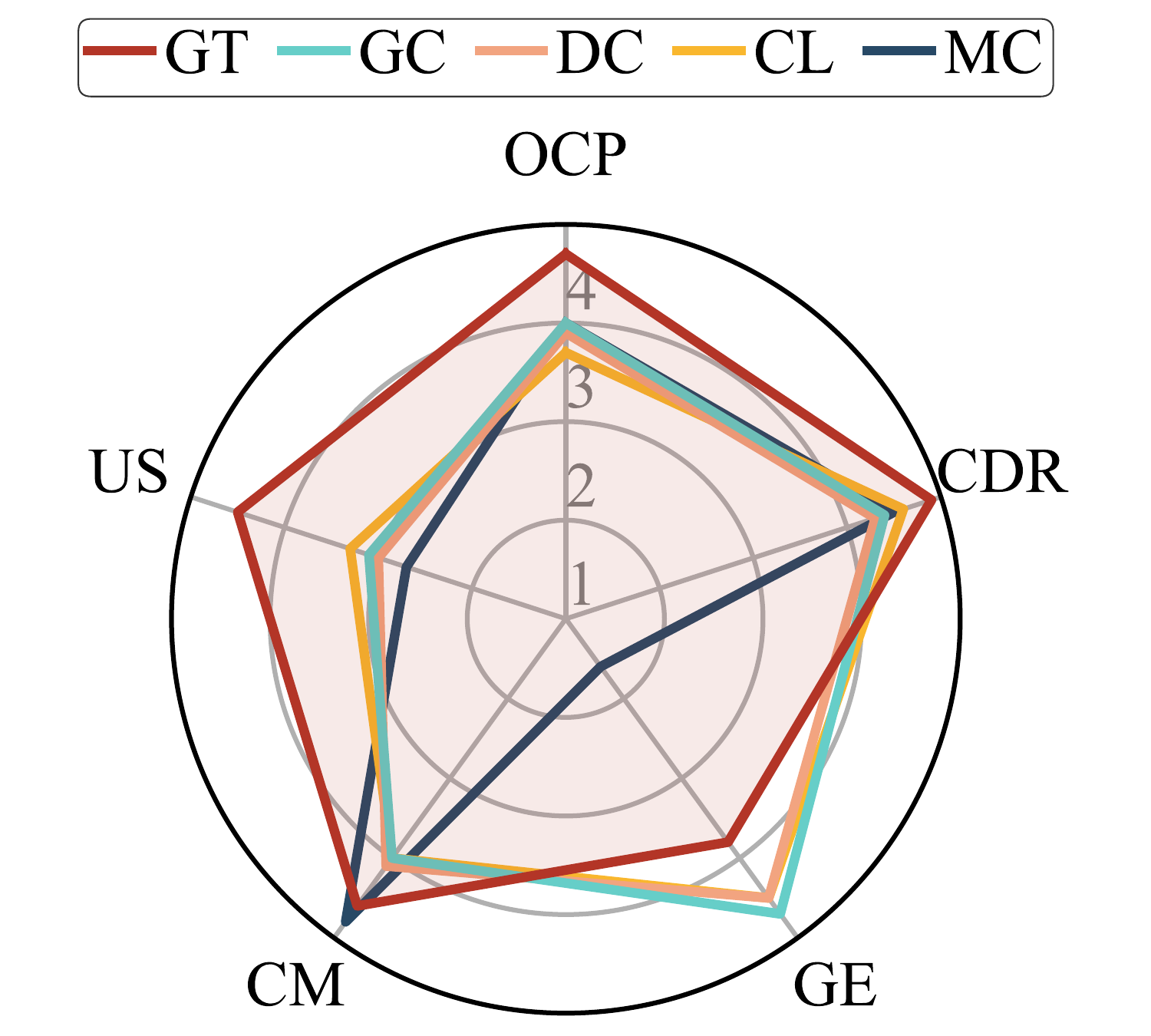}
        \label{fig:user_study_HAR}
    }
    \caption{User study on different tasks.}
    \label{fig:user_study}
    \vspace{-18pt}
\end{figure}

\vspace{-5pt}
\subsubsection{\textbf{RTSLM}}
\label{sec:requirement_transformation_ablation}
We directly feed the natural-language-described decomposed tasks from TDSLM into CGSLM. We then compare the performance of the generated code by \textit{GPIoT}, \textit{GPIoT} without RTSLM, and GPT-4o. As illustrated in Fig.~\ref{fig:ablation_requirement_transformation}, without RTSLM, we find that: 1) In HD, \textit{GPIoT} tends to generate code unrelated to the IoT domain with substantial precision degradation but a high recall rate. This implies that the heartbeat detection results contain numerous false positives.
2) In HAR, \textit{GPIoT} designs a basic HAR model with only a few simple layers, leading to a notable accuracy drop. These results highlight the importance of RTSLM in aiding CGSLM to understand the decomposed tasks generated by TDSLM, thereby improving its code generation ability for IoT applications.

\vspace{-8pt}
\subsection{User Study}
\label{sec:user_study}
\vspace{-2pt}
We conduct a user study to evaluate the functionality, generalizability, and overall satisfaction of \textit{GPIoT} for IoT application development. Specifically, with \textit{GPIoT} deployed on an edge server, we invite 5 experts and 15 non-experts in IoT and ask them to freely express their requirements for any IoT application development that requires signal processing or AI technologies. By sequentially executing the generated code based on the instructions in the documentation, we ask the users to rate \textit{GPIoT} based on five metrics: 1) \textit{Overall Code Performance} \textbf{(OCP)} evaluates the overall performance of the generated code on corresponding test data considering task accuracy, runtime efficiency, and resource consumption; 2) \textit{Code \& Documentation Readability} \textbf{(CDR)} measures the clarity and structure of the code and documentation; 3) \textit{Generation Efficiency} \textbf{(GE)} accesses how efficiently \textit{GPIoT} operates in terms of speed and resource usage to produce the final results; 4) \textit{Code Modularity} \textbf{(CM)} judges whether the code is properly modularized for easy reuse and extension; 5) \textit{User Satisfaction} \textbf{(US)} captures users' feedback regarding their overall personal experience. All the above metrics are rated by the users on a scale from 1 (not at all) to 5 (completely). Github Copilot, DeepSeek-Coder (cloud), CodeLlama (local), and MapCoder (agent) serve as representative baselines for comparison.

As shown in Fig.~\ref{fig:user_study}, we observe: 1) \textit{GPIoT} significantly outperforms the baselines in terms of OCP and US. The main reason is that, tuned on our IoT-specialized datasets, \textit{GPIoT} can generate code containing more dedicated algorithms with better performance. Therefore, the users provide a higher score for \textit{GPIoT} regarding the overall code performance and user satisfaction. 2) \textit{GPIoT} achieves similar scores to the baselines in terms of CDR and CM, because our datasets mainly focus on generating IoT-related code snippets. Therefore, the readability of the code and documentation are not explicitly enhanced via tuning. 3) \textit{GPIoT} gets a lower GE score as it performs requirement transformation and code generation for each decomposed task. Nevertheless, we can enhance its efficiency by adopting various LLM inference and serving optimization methods \cite{fu2024serverlessllm}. Additionally, during our implementation, we find that MapCoder costs around \$20 to synthesize a program for a single IoT application while \textit{GPIoT} incurs no query costs. The inferior performance of MapCoder may be attributed to the fact that LLM+RAG-based agents typically incorporate multiple modules to generate intermediate results in a cascaded manner, making them susceptible to unstable networks. This yields longer generation time, prohibitive token costs, and degraded user experiences. The user study results demonstrate the superior performance of \textit{GIoT} in synthesizing IoT-related programs in an end-to-end manner, with the ability to generalize to other unseen IoT applications.

\vspace{-8pt}
\section{Discussion}
\vspace{-2pt}
\textbf{System Cost of \textit{GPIoT}.}
In \textit{GPIoT}, though there are three SLMs (\ie, TDSLM, RTSLM, and CGSLM) operating simultaneously in \textit{GPIoT}, they share the same foundation model architecture and only differ in a small subset of tunable parameters (\S~\ref{sec:tuning}). Moreover, though LLM+RAG systems (\eg, MapCoder) require less memory (no more than 1 GB), \textit{GPIoT} does not need sophisticated prompt design and requires a shorter generation time for the final output (\S~\ref{sec:user_study}).

\noindent\textbf{Applicability to Resource-Constrained IoT Devices.} The main focus of \project is generating domain-specific and high-quality code for IoT application development. Considering the resource heterogeneity of various IoT devices for data processing, our \textit{IoT-Oriented Data Augmentation} method (\S~\ref{sec:augmentation}) augments the original task decomposition dataset by considering various resource requirements of different target platforms for IoT application development. Further experiments (Fig.~\ref{fig:resource_constrained_prompt}) demonstrate the effectiveness of \project in handling different resource requirements with optimized models.

\noindent\textbf{Generalizability of \textit{GPIoT}.} Existing IoT applications can be categorized into four types based on the functionality they deliver: data collection, data transmission, data processing, and decision-making. Data collection, transmission, and decision-making have been extensively studied using fixed programs. For instance, manufacturers typically provide COTS sample code for sensor data collection \cite{imu_collection}. In contrast, IoT data processing demands more complex algorithms due to fluctuating sensor data with noises and various resource constraints. Therefore, \project focuses on IoT data processing tasks by offering end-to-end solutions with executable programs. We believe that the general workflow (\ie, task decomposition $\rightarrow$ requirement transformation $\rightarrow$ code generation) of \project can be applied to other programming tasks. In future work, we plan to comprehensively assess \project in other complex programming tasks.

\vspace{-5pt}
\section{Related Work}

\vspace{-2pt}
\textbf{Code LLMs \& Programming Copilot.}
Code LLMs have significantly impacted the field of code generation, with prominent examples such as CodeLlama \cite{roziere2023code} and DeepSeek-Coder \cite{zhu2024deepseek}. Trained on vast datasets comprising diverse code repositories, these models are capable of synthesizing programs based on user requirements, effectively bridging the gap between natural language and code. One notable application is GitHub Copilot, an LLM-powered assistant that provides real-time code suggestions and auto-completion. Though powerful and promising, existing code LLMs and copilots are primarily designed for general-purpose programming, lacking customization to the IoT domain when tasked with IoT applications. \textit{GPIoT} addresses this by tuning local SLMs on IoT-specialized text-generation datasets with scrupulous augmentation. Additionally, by locally deploying \textit{GPIoT}, it can potentially serve as a copilot for IoT application developers, enhancing task accuracy and development efficiency in a privacy-preserving manner.


\noindent\textbf{Data Augmentation for LLMs.}
Existing text augmentation methods \cite{ding2024data, wang2024llm} for LLM tuning harness the advanced language processing capabilities of powerful LLMs (\eg, GPT-4) to synthesize diverse and high-quality text data. These methods have two categories: 1) Depth-based augmentation \cite{xu2023wizardlm, chen2024llm} aims to increase the complexity of the original text data by adding constraints, concretizing the problem, and increasing reasoning steps. 2) Breadth-based augmentation \cite{li2024data} directly uses powerful LLMs to rewrite the original text data and generate a completely new instruction. However, these augmentation methods focus on linguistic characteristics rather than the IoT domain knowledge. In \textit{GPIoT}, we propose a novel IoT-oriented text augmentation method tailored for the IoT domain, considering unique features of IoT applications, \ie, sensor modalities, data representations, and system resource constraints.



\vspace{-5pt}
\section{Conclusion}

\vspace{-2pt}
We present \textit{GPIoT}, a tailored local code generation system that synthesizes programs with documentation based on user requirements for IoT application development. Armed with two IoT-specialized text-generation datasets, the IoT-oriented augmentation method, and our PECT paradigm, \textit{GPIoT} can generate more IoT-related code in a privacy-preserving way, achieving enhanced task accuracy and user satisfaction for IoT application development. As IoT technologies are emerging rapidly, it is also worthwhile to explore the construction of a dynamic IoT knowledge database and continuous fine-tuning of local SLMs in the future.


\vspace{-5pt}
\begin{acks}
\vspace{-2pt}
We sincerely thank our anonymous shepherd and reviewers for their constructive comments and invaluable suggestions that helped improve this paper. This work is supported by Hong Kong GRF Grant No. 15211924 and 15206123. Yuanqing Zheng is the corresponding author.
\end{acks}

\clearpage
\balance
\bibliographystyle{ACM-Reference-Format}

\end{sloppypar}
\end{document}